%% file: main.tex
\documentclass[a4paper, amsfonts, amssymb, amsmath, reprint, showkeys, nofootinbib, twoside]{revtex4-1}
\usepackage[english]{babel}
\usepackage[utf8]{inputenc}
\usepackage[colorinlistoftodos, color=green!40, prependcaption]{todonotes}
\input{preamble}
\usepackage[pdftex, pdftitle={Article}, pdfauthor={Author}]{hyperref} % For hyperlinks in the PDF
\begin{document}
\title{Lead-free CsSnCl\textsubscript{3} perovskite nanocrystals: Rapid synthesis, experimental characterization and DFT simulations}

\author{Md. Shahjahan Ali, Subrata Das, Yasir Fatha Abed and M. A. Basith}
    \email[Email address: ]{mabasith@phy.buet.ac.bd}% Your name
    \affiliation{Nanotechnology Research Laboratory, Department of Physics, Bangladesh University of Engineering and Technology, Dhaka-1000, Bangladesh.\\ \\ DOI: \href{https://doi.org/10.1039/D1CP02666F}{10.1039/D1CP02666F}}

%\date{\today} % Leave empty to omit a date

\begin{abstract}
Lead-free metal halide perovskites have drawn great attention as light harvesters due to their promising optoelectronic and photovoltaic properties. In this investigation, we have successfully synthesized thermally stable cubic phase cesium tin chloride (CsSnCl\textsubscript{3}) perovskite nanocrystals (crystal size $\sim$300 nm) with improved surface morphology by adopting a rapid hot-injection technique. An excellent crystalline quality of these cubic shaped nanocrystals was confirmed by high-resolution transmission electron microscopy imaging. The binding of organic ligands on the surface of the sample was identified and characterized by the nuclear magnetic resonance spectroscopy. The UV-visible spectroscopy ensured that CsSnCl\textsubscript{3} nanocrystals have a direct band gap of $\sim$2.98 eV which was further confirmed by steady-state photoluminescence spectroscopy. The band edge positions calculated using the Mulliken electronegativity approach predicted the potential photocatalytic capability of the as-prepared nanocrystals which was then experimentally corroborated by the photodegradation of rhodamine-B dye under both visible and UV-visible irradiation. Our theoretical calculation by employing experimentally obtained structural parameters within generalized gradient approximation (GGA) and GGA + U methods demonstrated 90\% accurate estimation of experimentally observed optical band gap when U\textsubscript{eff} = 6 eV was considered. The ratio of the effective masses of the hole and electron expressed as D = $m_{h}^{*}/m_{e}^{*}$ was also calculated for U\textsubscript{eff} = 6 eV. Based on this theoretical calculation and experimental observation of the photocatalytic performance of CsSnCl\textsubscript{3} nanocrystals, we have proposed a new interpretation of the ``D'' value.  We think that a ``D'' value of either much smaller or much larger than 1 is the indication of low recombination rate of the photogenerated electron-hole pairs and the high photocatalytic efficiency of a photocatalyst. We believe that this comprehensive investigation might be helpful for the large-scale synthesis of thermally stable cubic CsSnCl\textsubscript{3} nanocrystals and also for a greater understanding of their potential in photocatalytic, photovoltaics and other prominent optoelectronic applications.
\end{abstract}

%\keywords{first keyword, second keyword, third keyword}

\maketitle

\input{sections/mainmanuscript.tex}  %I believe leaving the sections in separate files is more organized, change it if you desire 

\bibliography{PCCP} %You need to replace "rsc" on this line with the name of your .bib file
\bibliographystyle{rsc} %the RSC's .bst file

%\appendix*
%\clearpage
%\centering
%\input{sections/appendix1.tex}

\end{document}

%% file: preamble.tex
\usepackage{amsthm}
\usepackage{mathtools}
\usepackage{physics}
\usepackage{xcolor}
\usepackage{graphicx}
\usepackage[left=23mm,right=13mm,top=35mm,columnsep=15pt]{geometry} 
\usepackage{adjustbox}
\usepackage{placeins}
\usepackage[T1]{fontenc}
\usepackage{lipsum}
\usepackage{csquotes}

%% file: sections/mainmanuscript.tex
\section{Introduction} \label{sec:intro}
    Halide perovskites of ABX\textsubscript{3} stoichiometry where A is a monovalent organic (e.g., CH\textsubscript{3}NH\textsubscript{3}\textsuperscript{+}) or inorganic cation (e.g., Cs\textsuperscript{+} or Rb\textsuperscript{+}), B is a divalent metal ion (e.g. Sn\textsuperscript{2+}, Pb\textsuperscript{2+}, Mn\textsuperscript{2+} or Ge\textsuperscript{2+}) and X is a halogen (e.g. Cl, Br or I), have gained significant research focus over the past few years due to their promising optoelectronic applications, including semiconducting thin-film transistors, \cite{mao2018two,lin2017metal} light-emitting diodes (LED), \cite{yantara2015inorganic,xu2019rational,tian2016solution,ghosh2021nature} solar cells, \cite{pandey2019mutual,jung2017tio2,wang2021thickness,haris2021substance} photodetectors \cite{saleem2021cspbi3,wu2019air} and photovoltaics. \cite{teng2021first,sun2015chalcogenide,guo2021influence} Recently, a good number of all-inorganic metal halide perovskites (MHPs), especially cesium (Cs) based perovskites, CsBX\textsubscript{3} have gained a particular attraction due to their fascinating thermal stability \cite{jung2017influence} as well as intriguing optoelectronic properties. \cite{de2016highly, protesescu2015nanocrystals, liang2018enhancing, kang2018intrinsic} To date, various divalent metal ions such as Pb\textsuperscript{2+}, Sn\textsuperscript{2+}, Mn\textsuperscript{2+} as the B-site cation in CsBX\textsubscript{3} perovskites have been studied comprehensively and particularly, lead (Pb\textsuperscript{2+}) based halide perovskites have shown a record power conversion efficiency (PCE) of 25.5\% after years of research. \cite{jia2021unfused} However, from an environmental standpoint, the toxicity of Pb is still a frustrating issue for the large-scale production and industry-level applications of lead-based halide perovskites. Hence, from commercial prospects, recently a significant amount of research has been conducted by substituting Pb with non-toxic Sn to fabricate lead-free CsSnBr\textsubscript{3} and CsSnI\textsubscript{3} MHPs and thereby to investigate their potential applications in photodetectors, \cite{han2019controllable} LEDs, \cite{mu2020effects} LASERs, \cite{xing2016solution} X-rays, \cite{li2018all} memristors \cite{siddik2021nonvolatile} and synaptic devices. \cite{zhou2020perovskite} Compared to Br and I based MHPs, the experimental and theoretical investigations on the optical properties and electronic band structure of CsSnCl\textsubscript{3} perovskite are less explored. Particularly, the photocatalytic properties of CsSnCl\textsubscript{3} perovskite are yet to be investigated to the best of our knowledge. 

Notably, the fabrication of thermally stable CsSnCl\textsubscript{3} nanocrystals with desired morphology is quite challenging. A previous investigation reported a fabrication process of cubic CsSnCl\textsubscript{3} nanocrystals which required temperature-controlled mechanical milling and further 12 hour-long high temperature heat treatment. \cite{xia2020room} In a separate investigation, \cite{wu2019stabilizing} CsSnCl\textsubscript{3} perovskite of phases I and II were also synthesized successfully, however, the synthesis technique required high temperature processing steps. Apart from these, in a number of other investigations, \cite{yamada1998phase, peedikakkandy2016composition} CsSnCl\textsubscript{3} nanocrystals were synthesized by employing organic solvent-assisted treatment along with energy consuming high temperature annealing which are not suitable for time and cost-effective synthesis of halide perovskites. Therefore, a simple and fast alternative synthesis approach like hot-injection technique \cite{chen2016synthesis} is essential to fabricate CsSnCl\textsubscript{3} nanocrystals with desired structure, phase purity and surface morphology. Notably, the hot-injection technique has been widely employed to synthesize lead-free CsSnBr\textsubscript{3} and CsSnI\textsubscript{3} perovskite nanocrystals. \cite{wang2017controlled,kang2021antioxidative}

Furthermore, the experimental investigation of the material properties of halide perovskites at the molecular level are very complicated and sometimes unattainable due to the unavailability of the required experimental facilities. These complexities can be circumvented to a greater extent by implementing the density function theory (DFT) based first-principles calculation. By using the conventional DFT method, theoretical investigations of optical and electronic properties of CsSnCl\textsubscript{3} perovskite were carried out. \cite{idrissi2021band,korbel2016stability} In the aforementioned investigations, the theoretically calculated optical band gap of CsSnCl\textsubscript{3} perovskite was underestimated significantly if compared with the experimentally obtained band gap. \cite{idrissi2021band,korbel2016stability} In a previous investigation, \cite{korbel2016stability} computationally more costly hybrid exchange-correlation functional of Heyd, Scuseria, and Ernzerhof (HSE06) \cite{heyd2003hybrid} was used in DFT calculation. However, the theoretically calculated band gap of CsSnCl\textsubscript{3} deviated 45.9\% from the experimentally observed value. \cite{huang2013electronic} The accuracy of the optical properties and electronic band structure calculation of CsSnCl\textsubscript{3} is required to be improved for precise calculation of the optical and electronic band structure of CsSnCl\textsubscript{3} perovskite. In this regard, the effect of on-site Coulomb and exchange interaction might be considered within the generalized gradient approximation (GGA) by implementing the GGA+U method. \cite{Das2021,lin2021coherent,prasad2020synthesis} In GGA+U method the Hubbard U\textsubscript{eff} correction is implemented in a form of U\textsubscript{eff} = U – J, where the effects of the on-site Coulomb and exchange interaction are denoted by U and J, respectively. \cite{wang2006oxidation,dudarev1998electron}

Therefore, in this investigation, CsSnCl\textsubscript{3} perovskite was synthesized by hot-injection technique at temperature 200 $^{\circ}$C without performing any subsequent mechanical milling or any other energy-consuming high-temperature pre/post-processing.  We observed that the chemical compositions and optical band gap were in good agreement with other published results, \cite{siddik2021nonvolatile,peedikakkandy2016composition,huang2013electronic} which further confirmed the successful synthesis of CsSnCl\textsubscript{3} nanocrystals via the hot-injection method. Moreover, the synthesized CsSnCl\textsubscript{3} nanocrystals exhibited excellent absorption in the visible spectrum, which indicated its usability as a solar absorber and photocatalyst. Therefore, as a case study of its practical application, the photocatalytic dye degradation test of a well-known organic dye, rhodamine-B (RhB), was performed under both visible and UV-visible light irradiation. The photocatalytic dye degradation efficiency of up to 58\% was observed by irradiating solar light for 180 minutes using a solar simulator. Moreover, using both GGA and GGA+U methods, the effects of the on-site Coulomb and exchange interaction on absorption coefficient, band gap, band structure, the effective masses of charge carriers, charge density etc. have been investigated. The outcome of our theoretical calculations demonstrate that the GGA+U method incorporated with Perdew-Burke-Ernzerhof (PBE) functional provides reasonably accurate results close to the experimentally obtained values when U\textsubscript{eff} = 6 eV was considered. Through theoretical investigation, the ratio of the effective masses of electron  and hole was also evaluated to get a deep insight into the charge carrier recombination rate within the photocatalyst. 

\begin{figure*}[ht]
   \centering
   \includegraphics[width=0.9\textwidth]{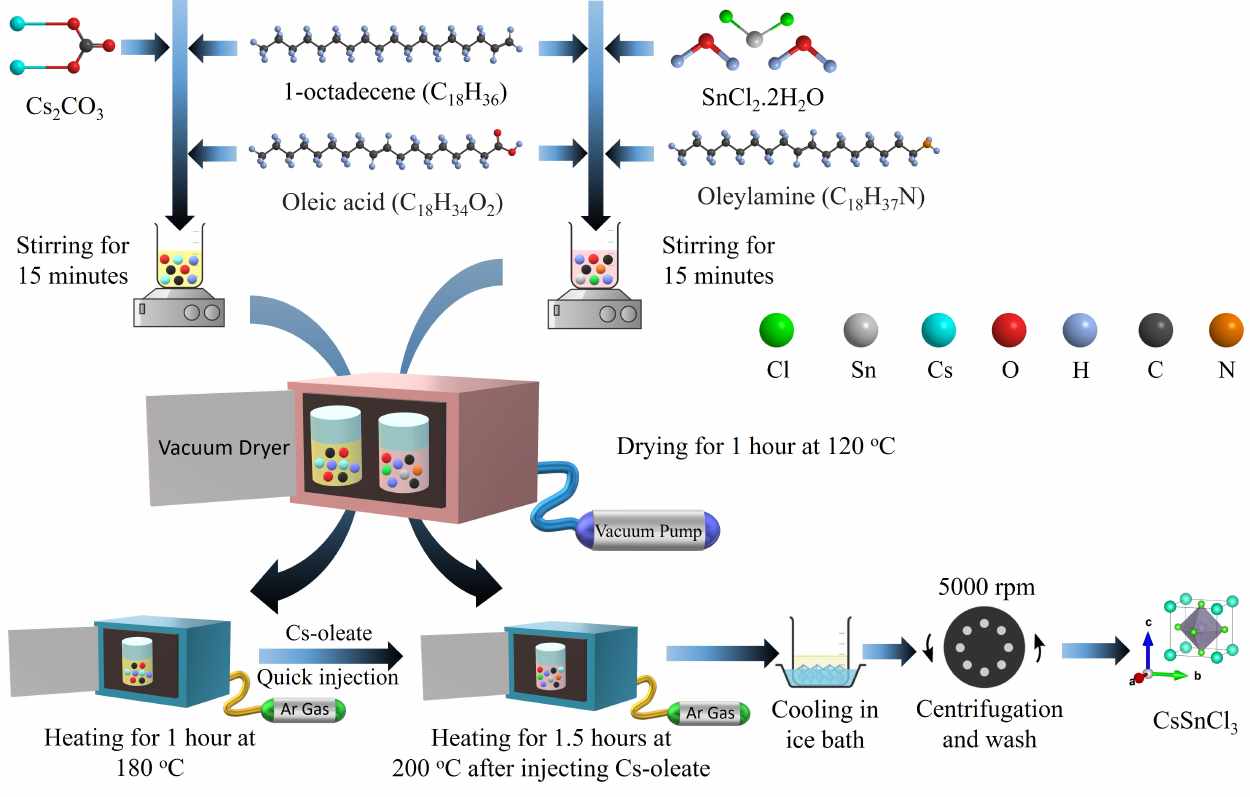}% Here is how to import EPS art
   \caption{\label{fig:syn} Schematically the synthesis steps of CsSnCl\textsubscript{3} perovskite using hot-injection technique.} 
\end{figure*}

\section{Experimental Details}

\subsection{Sample preparation}

In the present investigation, we have synthesized CsSnCl\textsubscript{3} perovskite by adopting a facile, low-temperature hot-injection technique. \cite{ murray1993synthesis, yu2002formation} Fig. \ref{fig:syn} depicts schematically the synthesis steps of CsSnCl\textsubscript{3} perovskite. At first, to prepare Cs-oleate, 0.40 mmol of Cs\textsubscript{2}CO\textsubscript{3} was added into the mixture of 15 mL of 1-octadecene and 1.25 mL of oleic acid. Then the mixture was rigorously stirred for 15 minutes and dried at 120 $^{\circ}$C for 1 hour in vacuum. Afterward, the resultant mixture was heated for 1 hour at 180 $^{\circ}$C in argon (Ar) atmosphere which led to the precipitation of the desired Cs-oleate. The obtained Cs-oleate precipitation was heated at 80 $^{\circ}$C for 30 minutes before using it for the synthesis of CsSnCl\textsubscript{3}. 

Next, for the synthesis of CsSnCl\textsubscript{3} perovskite, 225 mmol of SnCl$_2$.2H$_2$O, 30 mL of 1-octadecene, 3 mL of oleic acid and 3 mL of oleylamine were mixed together in a beaker by magnetic stirring and then dried at 120 $^{\circ}$C for 1 hour in vacuum. Afterward, 2.5 mL of Cs-oleate solution was injected rapidly into the beaker and then, the beaker was quickly transferred to a furnace working at inert atmosphere. Thereafter, the solution was heated at 200 $^{\circ}$C for 1.5 hours in Ar atmosphere. The mixture was then rapidly cooled using an ice bath. Finally, CsSnCl\textsubscript{3} nanocrystals were separated from the solution by centrifugation at 5000 rpm and were washed three times by n-hexane. Prior to further characterization, the obtained powder sample was dried at 80 $^{\circ}$C for 4 hours in vacuum. 

\subsection{Material characterization}
The crystal structure of the as-prepared sample was investigated by obtaining its powder X-ray diffraction (XRD) pattern using a diffractometer (Rigaku SmartLab) with a Cu X-ray source (wavelength, $\lambda$: K$_{\alpha 1}$ = 1.54059 \AA $\;$ and K$_{\alpha2}$ = 1.54180 \AA). Further, Rietveld refinement analysis of the XRD data was carried out by the FullProf computer program \cite{rodriguez1990fullprof} to determine the crystallographic parameters. Thermal analysis of the as-prepared CsSnCl\textsubscript{3} perovskite was performed by Thermogravimetric analysis (TGA) and Differential Scanning Calorimetry (DSC) measurement (NETZSCH, STA 449 F3 Jupiter) in N\textsubscript{2} atmosphere (heating rate 10 $^{\circ}$C/min). Fourier transform infrared (FTIR) spectrum of the sample was collected in the wavenumber range of 650 cm$^{-1}$ to 4000 cm$^{-1}$ using a Spectrum Two FT-IR Spectrometer (PerkinElmer). Further, the surface morphology of the sample was examined by a Field Emission Scanning Electron Microscope (FESEM) (XL30SFEG; Philips, Netherlands and S4300; HITACHI, Japan). In addition, elemental analysis was carried out by energy-dispersive X-ray (EDX) spectroscopy. Further, Tecnai G2 FE1 F12 transmission electron microscope (TEM) was used at an accelerating voltage of 200 kV to obtain the bright field images and selected area electron diffraction (SAED) pattern of the as-synthesized CsSnCl\textsubscript{3} perovskite. To determine the chemical binding energies and valence states of different compositional elements at the surface of the CsSnCl\textsubscript{3}, X-ray photoelectron spectroscopy (XPS) was conducted by a PHI5600 ESCA 120 (ULVAC PHI) system. The XPS spectrum was extracted using the monochromatized Al K$_\alpha$ X-ray (1486.6 eV) and 121 hemispherical mirror analyzer (total energy resolution $\sim$0.13 eV). The Shirley 123 procedure was applied to subtract the background of the XPS spectra and the Gaussian-Lorentzian function was used for peak fitting. For assessing the nuclear magnetic resonance (NMR) spectrum, a Bruker Avance III HD spectrometer was used with the \textsuperscript{1}H frequency of 400.17 MHz and temperature of 297.6 K. The spectral width was set to be as 14 ppm and 64k data points were sampled with a relaxation delay of 4 seconds. The absorbance spectrum of the material was obtained using an ultraviolet-visible (UV-visible) spectrophotometer (UV-2600, Shimadzu). The steady-state photoluminescence (PL) spectrum of the sample was obtained on an RF 6000 luminescence spectrometer made by Shimadzu, Japan. The photoluminescence lifetime of the sample was measured using a PicoQuant Fluotime 300 equipped with a PMA 175 detector and a LDH-P-C-440 diode laser with an excitation wavelength of 440 nm. The photocatalytic efficiency of the CsSnCl\textsubscript{3} nanocrystals was evaluated in the photodegradation of RhB dye under solar irradiation. The details of the photocatalytic experiments are provided in electronic supplementary information (ESI).

\section{Results and discussions}
\subsection{Experimental Analysis}
\subsubsection{Crystal structure analysis}
\begin{figure*}[t]
   \centering
   \includegraphics[width=15 cm]{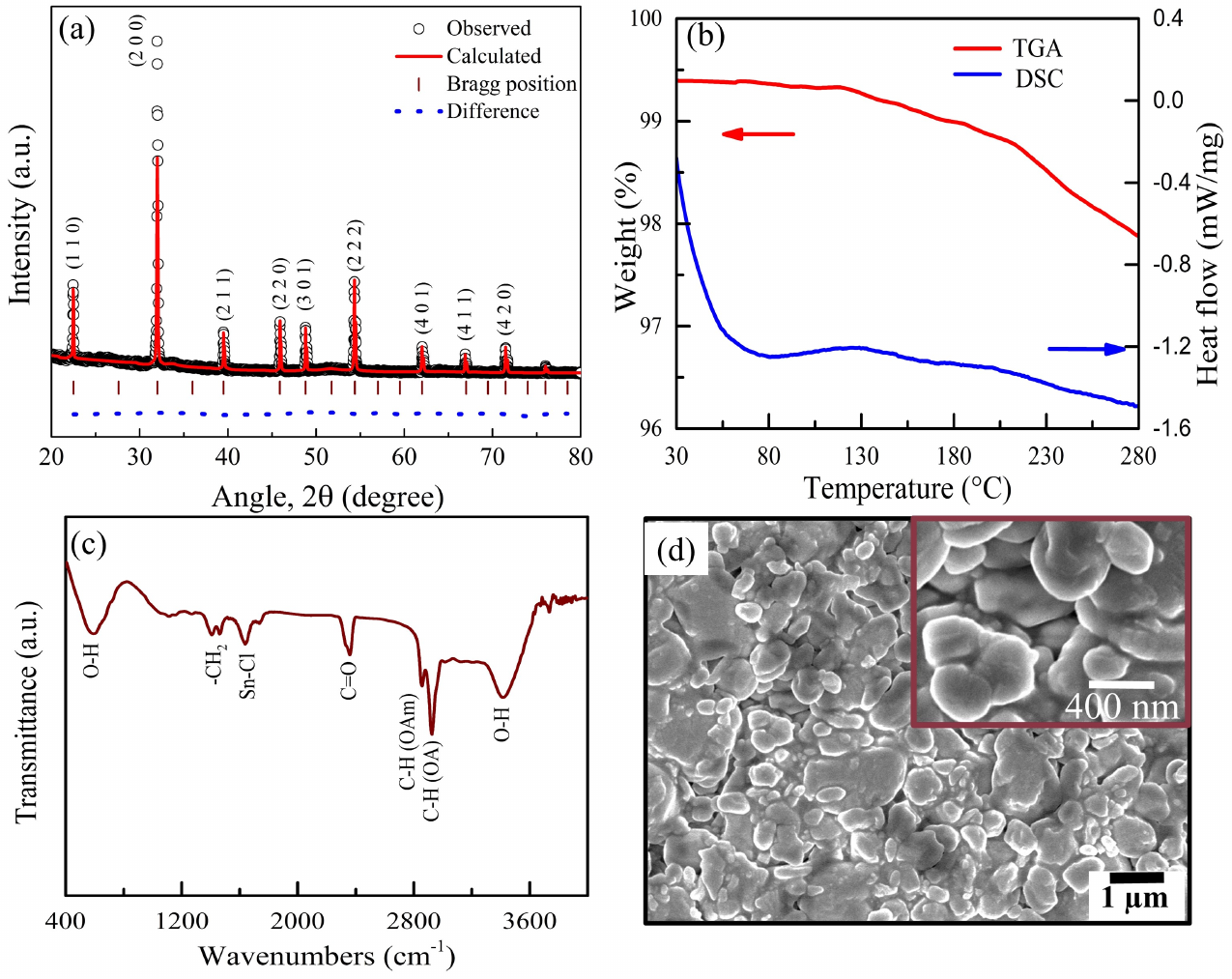}% Here is how to import EPS art
   \caption{\label{fig:xrd_to_FESEM} (a) Rietveld refined powder XRD patterns of CsSnCl\textsubscript{3} perovskite. (b) TGA and DSC curves of cubic CsSnCl\textsubscript{3} perovskite. (c) FTIR spectrum recorded at room temperature. (d) FESEM image of CsSnCl\textsubscript{3} perovskite nanocrystals. Inset: A magnified view of the as-synthesized nanocrystals.} 
\end{figure*}

The crystallographic structure and phase composition of the as-synthesized CsSnCl\textsubscript{3} perovskite were investigated by analyzing its Rietveld refined powder XRD spectrum. As shown in Fig.  \ref{fig:xrd_to_FESEM} (a), the Bragg reflected peaks exhibited by the sample can be indexed at (110), (200), (211), (220), (301), (222), (401), (411) and (420) planes
(JCPDS card file no 22-0200) which conform to the cubic crystal phase (\emph{Pm3m} space group) of CsSnCl\textsubscript{3} perovskite. \cite{xia2020room} Notably, the XRD pattern of as-synthesized CsSnCl\textsubscript{3} did not exhibit any undesirable peaks which indicates its high-phase purity. By Rietveld refinement analysis, the lattice parameters of as-synthesized CsSnCl\textsubscript{3} perovskite were determined as $a$ = $b$ = $c$ = 5.583 \AA, and $\alpha$ = $\beta$ = $\gamma$ = 90$^{\circ}$ with unit cell volume of 174.02 \AA$^3$. For a further understanding, the atomic positions and bond length of CsSnCl\textsubscript{3} as well as the reliability (R) factors of Rietveld refinement were inserted in Table 1.

The thermal stability of as-synthesized CsSnCl\textsubscript{3} sample was investigated by conducting TGA and DSC measurements. The outcomes of these measurements are presented in Fig. \ref{fig:xrd_to_FESEM} (b). As can be observed from the TGA curve, the weight loss of CsSnCl\textsubscript{3} perovskite due to the increase of temperature from 30 $^{\circ}$C to 280 $^{\circ}$C was only 2.15\% which is an indication of its excellent thermal stability. Such a small percentage of weight loss indicates the presence of nominal amount of binding surface ligands i.e. oleylamine and oleic acid on the as-synthesized nanocrystals. \cite{de2016highly, dastidar2017quantitative, hoffman2017cspbbr3}

Further, as illustrated in Fig.  \ref{fig:xrd_to_FESEM} (b), no endothermic or exothermic peak was observed in the DSC curve of synthesized CsSnCl\textsubscript{3} perovskite indicating that any crystallographic phase transition did not occur in our fabricated sample within the temperature range of 30 $^{\circ}$C to 280 $^{\circ}$C. The observed initial decrease in the DSC curve can be attributed to the moisture removal from the sample. Here, it is intriguing to note that according to a recent investigation, \cite{xia2020room} CsSnCl\textsubscript{3} crystals synthesized by mechanical milling technique undergoes monoclinic to cubic phase transition at around 120 $^{\circ}$C and hence, heat treatment at 120 $^{\circ}$C is required to obtain stable cubic phase using this technique . In contrast, the as-synthesized cubic-phase CsSnCl\textsubscript{3} perovskite has demonstrated high structural stability over a wide temperature range which eliminates the necessity of post-processing heat treatment and  justifies the significance of employing rapid hot-injection technique to synthesize CsSnCl\textsubscript{3} with a stable crystallographic phase.

\begin{table*}
    \centering
    \small
    \caption{Structural parameters of CsSnCl\textsubscript{3} nanocrystals and the value of R factors.}
    %\resizebox{\textwidth}{!}{
    \begin{tabular}{c c c c c c c c c c c}
    \hline 
    Atom & Wyc. Positions & $x$ & $y$ & $z$ & $a$ (\AA) & $\alpha$ & volume (\AA \textsuperscript{3}) & Bond length (\AA) & R factors \\
    \hline
     Cs & 1a & 0.0 & 0.0 & 0.0 & &  & & & R$_{exp}$=5.48\\
     Sn & 1b & 0.5 & 0.5 & 0.5 &5.583 & 90$^{\circ}$ & 174.02  & Sn-Cl = 2.791 & R$_{wp}$=7.24\\ 
     Cl & 3c & 0.0 & 0.5 & 0.5 & & & & & $\chi^2$=2.53\\
    \hline
    \end{tabular}%}
    \label{tab:riet}
\end{table*}

\subsubsection{FTIR analysis}

Further, FTIR spectroscopy was conducted to identify the presence of different functional groups within the CsSnCl\textsubscript{3} perovskite. As shown in Fig.  \ref{fig:xrd_to_FESEM} (c), the FTIR spectrum of as-prepared CsSnCl\textsubscript{3} nanocrystals exhibited a broad band at 590 cm\textsuperscript{-1} which can be attributed to the in-plane O-H group bending vibration. \cite{degen1997tables, feng2009anodization} The two weak absorbance peaks observed at 1406 and 1462 cm\textsuperscript{-1} are ascribed to the CH$_2$ bending. \cite{delgado2018thermal} The absorption band at 1637 cm\textsuperscript{-1} might have arisen due to the Sn-Cl bond formation. \cite{feng2009anodization} The discerned peak around 2360 cm\textsuperscript{-1} can be associated with the C=O vibrations. \cite{hills2017synthesis} The vibrational bands at 2856 and 2925 cm\textsuperscript{-1} have emerged due to the C-H stretching vibrations at the methylene groups of the precursors i.e. oleylamine and oleic acid, respectively. \cite{kim2021coordinating, morales2013effect} In addition, the broad absorption band at 3417 cm\textsuperscript{-1} can be imputed to the stretching vibration of O-H functional group. Thus, the outcomes of FTIR analysis conspicuously justify the successful formation of phase-pure CsSnCl\textsubscript{3} perovskite as was also evident from the XRD analysis.

\subsubsection{Morphological and elemental analysis}

The FESEM image of CsSnCl\textsubscript{3} perovskite is presented in Fig.  \ref{fig:xrd_to_FESEM} (d) which demonstrates that the surface of our as-prepared sample was satisfactorily homogeneous and non-porous. Noticeably, in a previous investigation, agglomeration in CsSnCl\textsubscript{3} perovskite was observed although the material was synthesized by mechanical milling under inert atmosphere and subsequent heat treatment for 12 hours. \cite{xia2020room} On the contrary, the hot-injection technique employed in this investigation yielded non-agglomerated CsSnCl\textsubscript{3} perovskite crystals successfully as shown in Fig.  \ref{fig:xrd_to_FESEM} (d). For a better insight, we have presented a magnified view of as-synthesized sample in the inset of Fig.  \ref{fig:xrd_to_FESEM} (d). The Fig. S1 inserted in the ESI shows the distribution of particle size. From these two figures, it can be estimated that the average particle size lies in the sub-micrometer region which confirms the successful synthesis of CsSnCl\textsubscript{3} nanocrystals. Further, the elemental composition of as-prepared CsSnCl\textsubscript{3} nanocrystals was assessed from EDX analysis at room temperature. The mass and atom percentages of each element in CsSnCl\textsubscript{3} as obtained by EDX are provided in ESI Table S1. For comparison, we have also tabulated the theoretically calculated values in ESI Table S1. Clearly, the mass and atom percentages of the desired elements i.e. Cs, Sn and Cl in CsSnCl\textsubscript{3} nanocrystals are in good agreement with the theoretical values which provides further evidence for the successful synthesis of  CsSnCl\textsubscript{3} perovskite.

\begin{figure*}[t]
   \centering
   \includegraphics[width=0.7\textwidth]{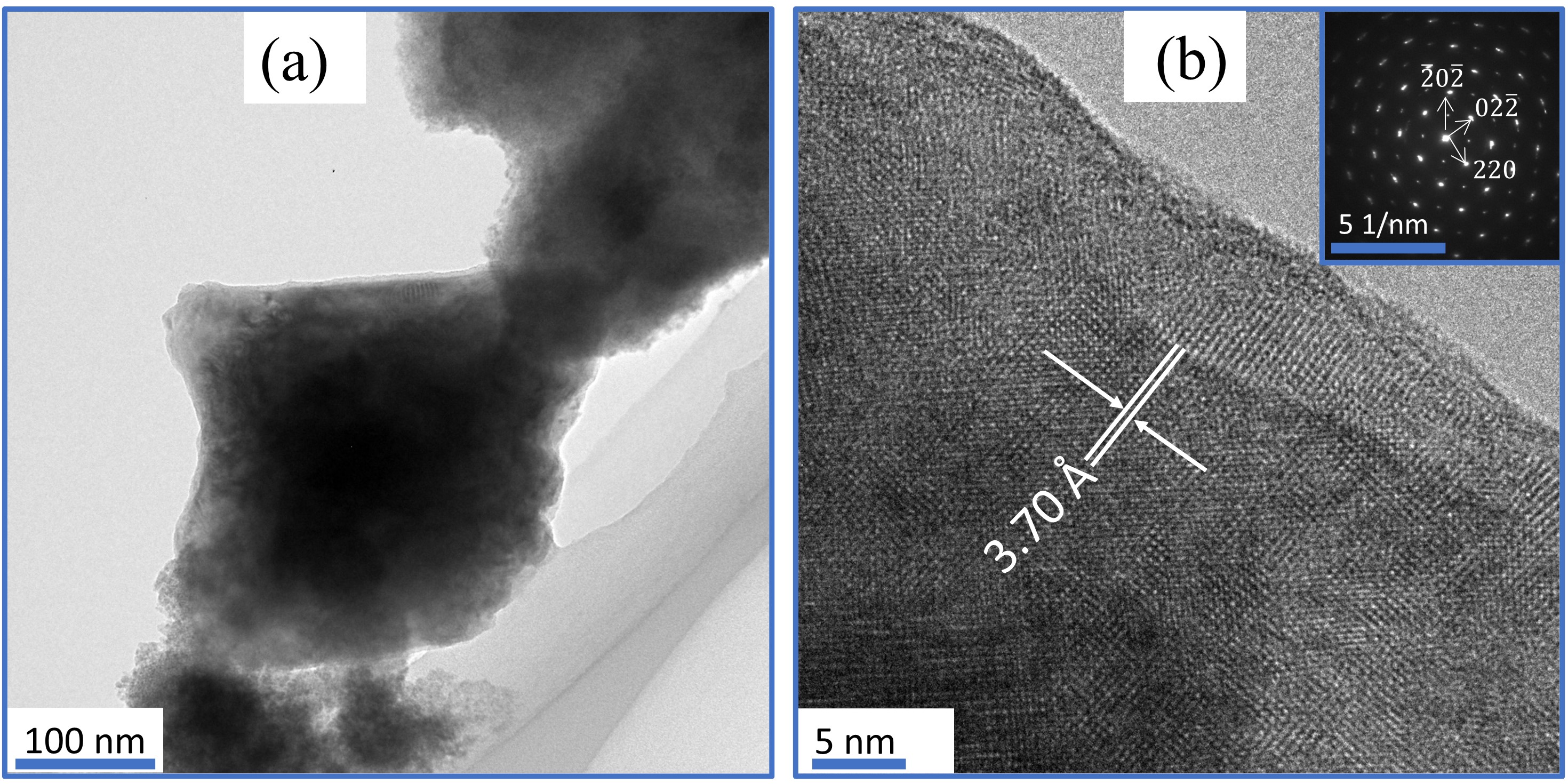}% Here is how to import EPS art
   \caption{\label{fig:TEM} (a) Bright field TEM image of synthesized CsSnCl\textsubscript{3} perovskite nanocrystals. (b) HRTEM image of CsSnCl\textsubscript{3} perovskite. Inset: SAED pattern of CsSnCl\textsubscript{3} nanocrystals.}
\end{figure*}

Fig. \ref{fig:TEM} (a) demonstrates the bright field TEM image of CsSnCl\textsubscript{3} nanocrystal which is of cubic shape. The high resolution TEM (HRTEM) image of Fig. \ref{fig:TEM} (b) shows the crystallinity of this nanocrystal with a 3.70 {\AA} interplanar spacing. The inset of Fig. \ref{fig:TEM} (b) reveals the selected area electron diffraction (SAED) pattern of the CsSnCl\textsubscript{3} nanocrystal. The SAED pattern demonstrates the presence of (202), (022), and (220) planes of the face centered cubic phase which further ensures the formation of the cubic structure of this nanocrystal. \cite{jellicoe2016synthesis, lin2019phase}

\begin{figure*}[t]
   \centering
   \includegraphics[width=15 cm]{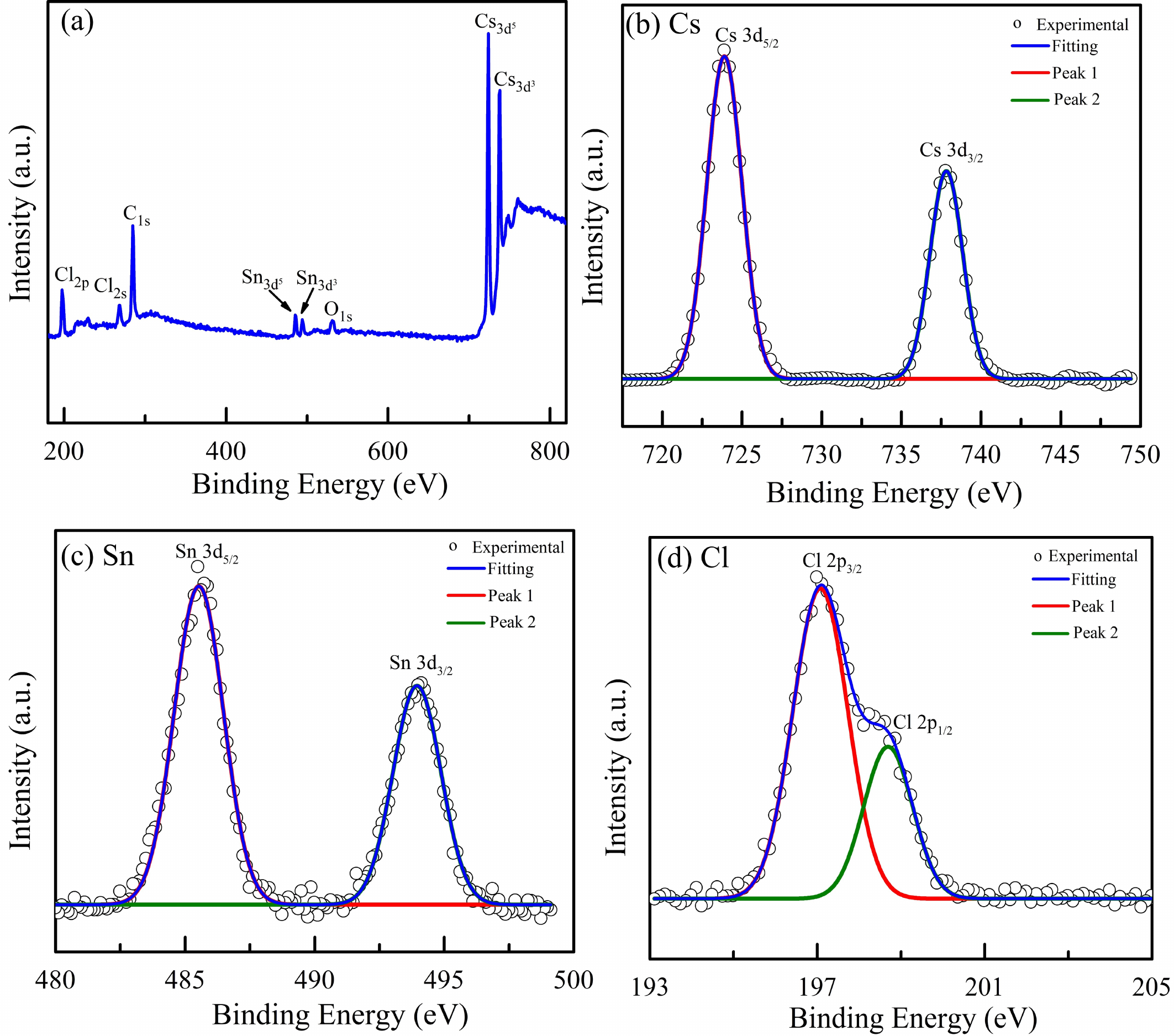}% Here is how to import EPS art
   \caption{\label{fig:XPS} (a) XPS full spectra of CsSnCl\textsubscript{3} nanocrystals  illustrating the existence of constituent elements (Cs, Sn and Cl) in the fabricated sample. Core level XPS spectra for (b) Cs 3d (c) Sn 3d (d) Cl 2p demonstrating the purity of the sample and valence states of the constituent elements, respectively.} 
\end{figure*}

\subsubsection{Chemical state analysis}

The chemical composition of as-prepared CsSnCl\textsubscript{3} nanocrystals was investigated by performing XPS analysis and the obtained full survey XPS spectra has been demonstrated in Fig. \ref{fig:XPS} (a). In this survey spectra, strong peaks for Cs, Sn and Cl core levels were identified which confirms the successful synthesis of CsSnCl\textsubscript{3} nanocrystals with the desired surface chemical states. Notably, two additional peaks corresponding to C 1s and O 1s orbitals were also noticed in the XPS spectrum of CsSnCl\textsubscript{3} [Fig. \ref{fig:XPS} (a)]. The emergence of these peaks may be due to the molecular adsorption of oxygen and carbon on the surface of as-prepared CsSnCl\textsubscript{3} nanocrystals during the exposure to the atmosphere which is common in the XPS investigation of related perovskites. \cite{musazay2015experimental, jellicoe2016synthesis} For a further insight, we have presented the high resolution XPS core spectra of Cs 3d, Sn 3d and Cl 2p orbitals in Fig. \ref{fig:XPS} (b)-(d), respectively. In Fig. \ref{fig:XPS} (b), two distinct peaks can be observed at the binding energies of 723.9 eV and 737.8 eV in the Cs 3d XPS spectrum which can be ascribed to the presence of Cs 3d\textsubscript{5/2} and Cs 3d\textsubscript{3/2} states, respectively. \cite{ zhang2019enhanced}  As shown in Fig. \ref{fig:XPS} (c), the core level spectrum of Sn was distinguished by the two pronounced peaks at 485.6 eV and 494 eV corresponding to the states of Sn 3d\textsubscript{5/2} and Sn 3d\textsubscript{3/2}, respectively. \cite{allioux2020bi} In Fig. \ref{fig:XPS} (d), the Cl 2p core level of CsSnCl\textsubscript{3} exhibits an asymmetric peak which has been de-convoluted into two symmetric Gaussian peaks at 197.1 eV and 198.5 eV. Notably, the peak at the lower binding energy, i.e. 197.1 eV is associated with the Cl 2p\textsubscript{3/2} state, while the higher binding energy peak can be attributed to the existence of Cl 2p\textsubscript{1/2} state in as-prepared CsSnCl\textsubscript{3}  nanocrystals. \cite{bera2017flexible, cheng2012enhanced}

Further, to identify the binding of surface ligands of the CsSnCl\textsubscript{3} perovskite nanocrystals, we have performed the one-dimensional (1D) \textsuperscript{1}H NMR spectroscopy. Prior to the characterization, the CsSnCl\textsubscript{3} nanocrystals were disseminated in deuterated CDCl\textsubscript{3} solvent. The 1D \textsuperscript{1}H NMR spectrum of the nanocrystals obtained under the aforementioned condition are demonstrated in the ESI Fig. S2. Interestingly, the peak formation is comparable with that of analogous perovskite nanocrystals. However, an up-field chemical shift in the peaks can be detected which might be attributed to the electronegativity of the Cl atoms. \cite{de2016highly} For deeper speculation, we have compared the obtained NMR spectrum with previously reported NMR spectra of oleic acid, oleylamine, and octadecene. \cite{de2016highly, ravi2017origin} It can be clearly seen that the peaks below 2.2 ppm appear because of the oleic acid, octadecene, and oleylamine ligands, but as their peaks are overlapping owing to their chemical shift, it is difficult to separate their individual peaks in this regime. Around 2.44 ppm, we found a relatively weak peak which is the characteristic peak of the oleic acid. Evidently, at around 3.25 ppm, a very wide peak is obtained, which can be attributed to the CH\textsubscript{2}-N and similarly, around 6.87 ppm, we have obtained another broad peak which corresponds to the $\text{NH}_3^+$ originated from the oleyammonium. These two peaks are unambiguously correlated to the oleylamine, which is also in well agreement with the previous reports. \cite{de2016highly} Notably, the peak at 5.35 ppm is associated with the alkane proton, which might have originated from either the oleylamine or oleic acid or from both of them. Such observation eventually conforms to the presence of protonated oleylamine. \cite {de2016highly} Interestingly, we did not observe the characteristic resonances of a terminal alkene presumably octadecene which provides evidence for the surface purity of the as-synthesized CsSnCl\textsubscript{3} nanocrystals unlike CsPbBr\textsubscript{3} nanocrystals reported in a previous investigation.\cite{de2016highly} 

\begin{figure*}[t]
   \centering
   \includegraphics[width=14 cm]{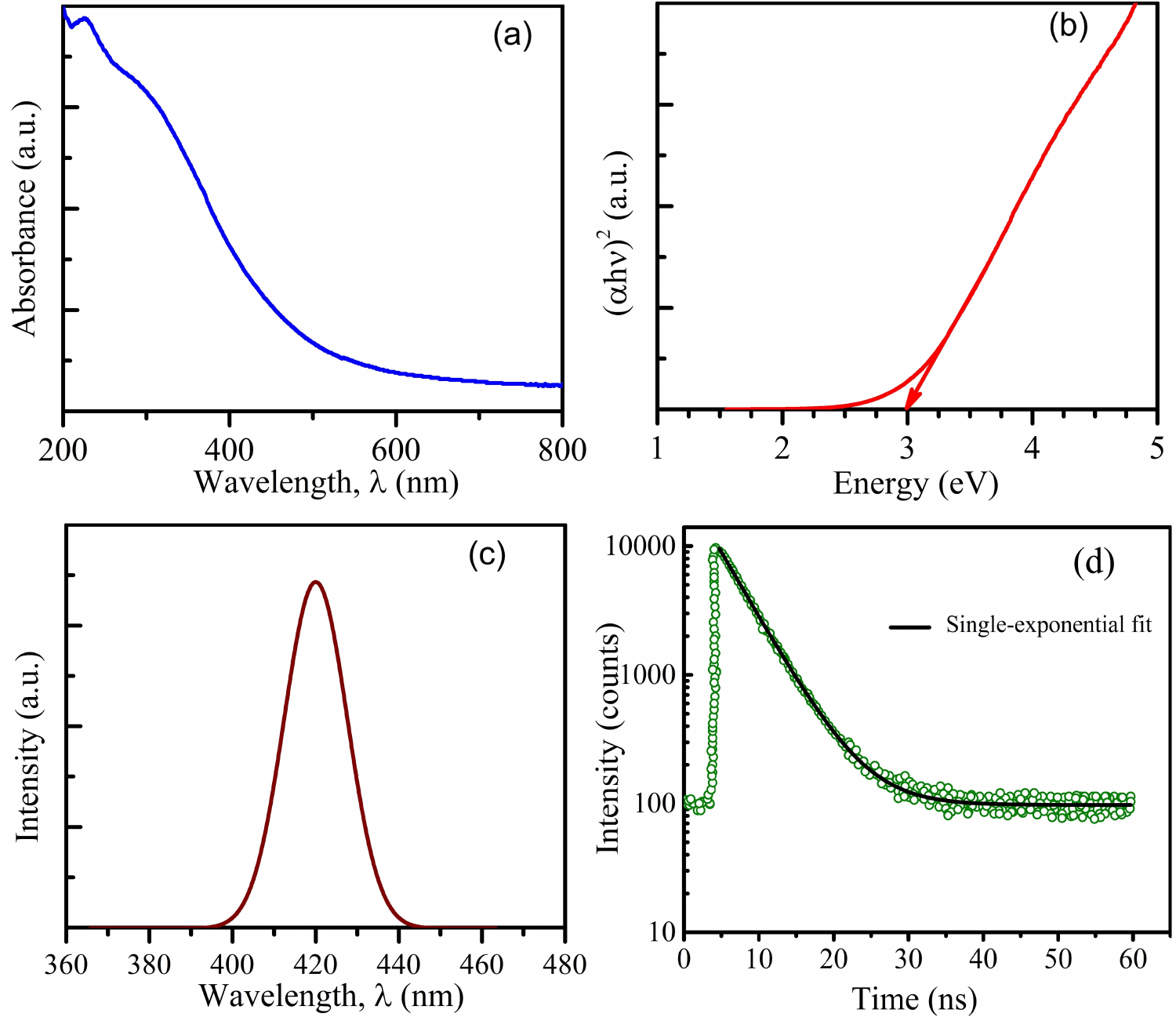}% Here is how to import EPS art
   \caption{\label{fig:optic} (a) Absorption spectrum of CsSnCl\textsubscript{3} nanocrystals. (b) Tauc plot of CsSnCl\textsubscript{3} perovskite demonstrating its direct band gap as $\sim$2.98 eV. (c) Steady-state PL spectrum of CsSnCl\textsubscript{3}. (d) Time resolved PL decay curve under pulsed 440 nm excitation at room temperature for CsSnCl\textsubscript{3} nanocrystals.} 
\end{figure*}

\subsubsection{Experimentally obtained optical properties}

The optical properties of the CsSnCl\textsubscript{3} nanocrystals were studied by conducting UV-visible absorption spectroscopy.  From Fig.  \ref{fig:optic}(a), it can be seen that the CsSnCl\textsubscript{3} nanocrystals show sufficiently higher absorption in the visible region as compared to the infrared region. Moreover, a strong absorption peak is noticeable in the ultraviolet region of the absorption spectrum of CsSnCl\textsubscript{3} which suggests its applicability for solar energy harvesting. Further, we employed the absorbance data to determine the optical band gap of our as-synthesized CsSnCl\textsubscript{3} nanocrystals using the Tauc relation. \cite{tauc1966optical} Since a number of previous investigations have reported the band gap of CsSnCl\textsubscript{3} to be direct, \cite{huang2013electronic, zhang2019disappeared} hence we have used the Tauc relation for direct band gap materials as shown in Fig. \ref{fig:optic}(b). From the Tauc plot, the direct optical band gap of as-synthesized CsSnCl\textsubscript{3} can be estimated as $\sim$2.98 eV which is well consistent with the literature. \cite{siddik2021nonvolatile,peedikakkandy2016composition,huang2013electronic} We have also measured the PL spectra of CsSnCl\textsubscript{3} to have an insight into the charge carrier recombination phenomenon. The PL peak shown in Fig. \ref{fig:optic}(c) is an indication of radiative recombination of photoexcited electron-hole (e$^{-}$-h$^{+}$) pairs. The direct band gap value can also be estimated from the PL peak as $\sim$2.96 eV which agrees well with the outcome of the Tauc plot (Fig. \ref{fig:optic}(b)).\\

Further, in order to measure the excited state lifetime of the charge carriers in the CsSnCl\textsubscript{3} nanocrystals, we have employed the time-correlated single photon
counting (TCSPC) technique and presented the time-resolved PL decay curve in Fig. \ref{fig:optic} (d). The average lifetime ($\tau _{av}$) of the charge carriers was calculated as 4.27 ns by fitting the decay curve with the single-exponential function. Notably, the photoluminescence quantum efficiency (PLQY) of the sample was determined to be 91\%. Radiative lifetime was also calculated from $\tau _{av}$ and PLQY using equation (1). The non-radiative lifetime ($\tau _{nr}$) and the radiative ($k_{r}$) /non-radiative ($k_{nr}$) recombination rate constants were finally obtained using equation (2) and (3). Notably, the calculated values of $k_{r}$ and $k_{nr}$ of as-synthesized CsSnCl\textsubscript{3} nanocrystals were 0.213 ns\textsuperscript{-1} and 0.021 ns\textsuperscript{-1}, respectively which are comparable with related previous investigations. \cite{jellicoe2016synthesis,koscher2017essentially,ghosh2019influence} \\

\begin{equation}
    PLQY =\frac{\tau _{av}}{\tau _{_{r}}}
    \end{equation}
\begin{equation}
     \frac{1}{\tau _{av}}=\frac{1}{\tau _{r}}+\frac{1}{\tau _{nr}} 
\end{equation}
  \begin{equation}
          k_{r}=\frac{1}{\tau _{r}},  k_{nr}=\frac{1}{\tau _{nr}}
  \end{equation}

In order to assess the potential of CsSnCl\textsubscript{3} nanocrystals for photocatalytic applications, we have determined their band edge positions by adopting the Mulliken electronegativity approach. The conduction band minimum (CBM) and valence band maximum (VBM) potential (vs. normal hydrogen electrode potential, NHE) of CsSnCl\textsubscript{3} were calculated as -0.44 and 2.54 eV, respectively. An energy band diagram is provided in ESI Fig. S3 consolidating the obtained values of CBM and VBM along with the redox potentials of different redox half-reactions of our interest. According to thermodynamics, to act as an effective catalyst of photocatalytic dye degradation, a material must possess a VBM $>$ 2.38 eV and CBM $<$ -0.16 V for executing reduction and oxidation half-reactions, respectively. \cite{tama2019mos} Clearly, both the CBM and VBM potentials of as-synthesized CsSnCl\textsubscript{3} nanocrystals meet the aforementioned requirements which indicate their promising potential for the photocatalytic degradation of different organic dyes. Moreover, the calculated band edge positions of CsSnCl\textsubscript{3} nanocrystals shown in ESI Fig. S3 are also favorable for photocatalytic hydrogen evolution via water splitting. Hence, further, we had assessed the photocatalytic efficiency of CsSnCl\textsubscript{3} nanocrystals which will be discussed in the next section.

\subsubsection{Photocatalytic degradation activity}

\begin{figure*}[t]
   \centering
   \includegraphics[width=15 cm]{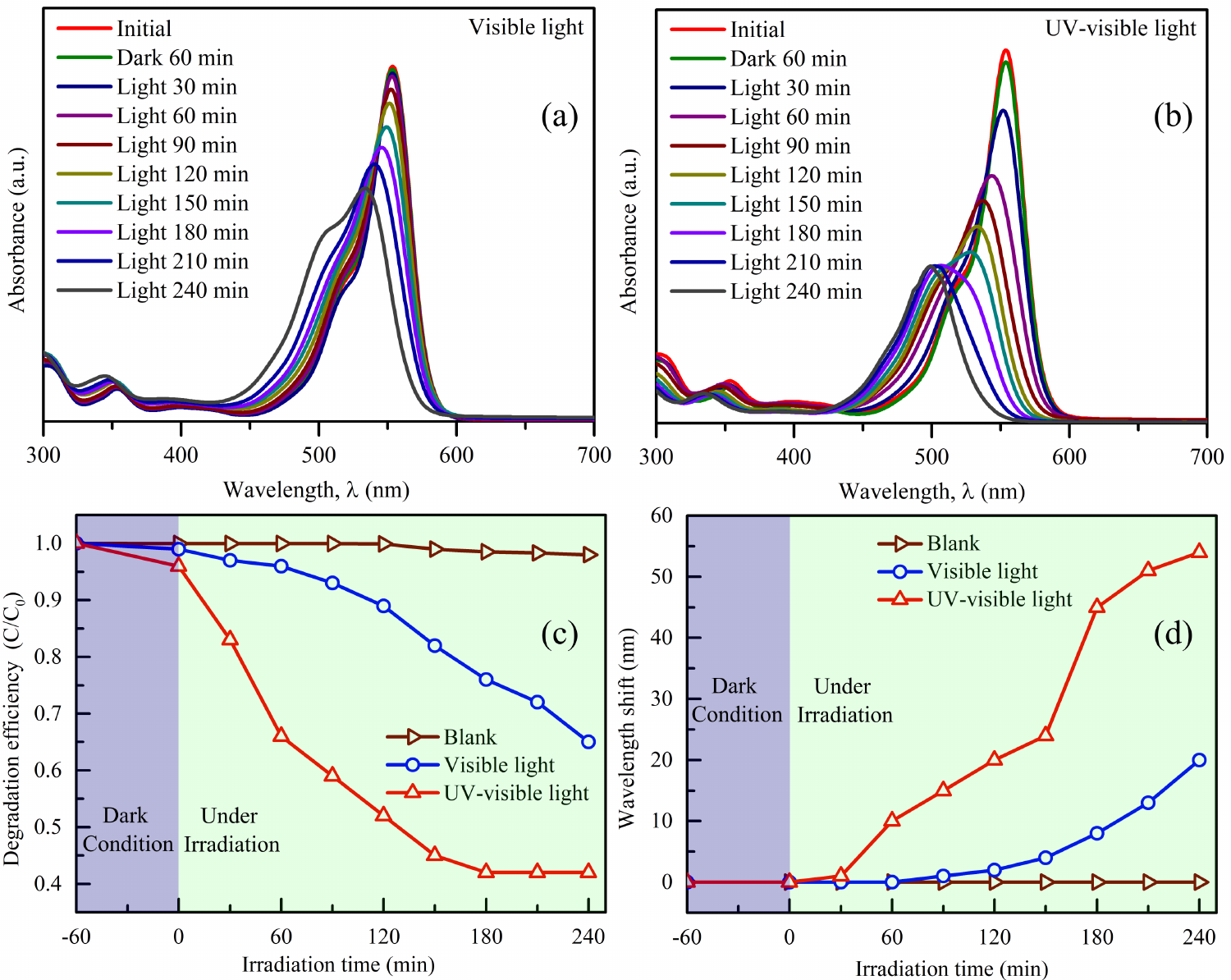}% Here is how to import EPS art
   \caption{\label{fig:photocatalytic} Time-dependent absorption spectra of RhB solution for different times up to 240 minutes under the irradiation of the (a) visible and (b) UV-visible light. (c) Degradation efficiency of RhB as a function of the irradiation time for CsSnCl\textsubscript{3} photocatalyst. (d) The wavelength shifts of the absorption spectra  of the RhB dye solution under identical conditions of figure (c). } 
\end{figure*}

The photocatalytic performance of the CsSnCl\textsubscript{3} nanocrystals was investigated as a case study towards the degradation of RhB under both visible and UV-visible illumination. Initially, for both experimental conditions, a typical RhB blank test and dark-adsorption test were carried out. The outcome of these tests confirmed the negligible self-photolysis and chemi-adsorption potential of the RhB dye molecules implying that the decomposition of RhB is solely attributed to photocatalysis. Fig. \ref{fig:photocatalytic}(a) and \ref{fig:photocatalytic}(b) demonstrate the change in the absorbance spectra of RhB photodegraded over as-synthesized CsSnCl\textsubscript{3} nanocrystals under the irradiation of visible and UV-visible light, respectively.  Clearly, under visible light illumination (Fig. \ref{fig:photocatalytic}(a)), the intensity of the characteristic absorption peak at 553 nm decreased gradually up to 240 minutes indicating the successful decomposition of RhB. On the other hand, for UV-visible irradiation (Fig. \ref{fig:photocatalytic}(b)), the peak decreased more rapidly up to 180 minutes but no significant degradation could be observed after that. Fig. \ref{fig:photocatalytic}(c) provides the degradation efficiency vs. irradiation time curves where \textit{C\textsubscript{0}} and \textit{C} denote the initial and residual concentrations measured at 30 minutes of interval, respectively. It can be seen that CsSnCl\textsubscript{3} degraded $\sim$35\% of RhB dye after 240 minutes of visible light irradiation. Interestingly, the degradation percentage of RhB increased to $\sim$58\% within only 180 minutes under UV-visible irradiation. Such enhancement in photocatalytic efficiency of as-synthesized CsSnCl\textsubscript{3} nanocrystals can be attributed to their reasonably wide-band gap value, $\sim$2.98 eV (Fig. \ref{fig:optic}(d)) along with good crystallinity and excellent surface morphology. It can be anticipated that the photoexcitation of charge carriers was enhanced significantly under the illumination of high energy UV photons which eventually resulted in a higher RhB dye degradation rate.

It is worth noting that RhB dye typically degrades via two competitive and co-existing pathways i.e. the cleavage of its whole conjugated chromophore structure and N-deethylation process. \cite{watanabe1977photocatalysis,hu2006oxidative} In the first pathway, the absorption peak intensity decreases while the peak position remains constant. On contrary, during N-deethylation, the characteristic peak gradually shifts to shorter wavelengths with constant intensity. Therefore, in Fig. \ref{fig:photocatalytic}(a) and \ref{fig:photocatalytic}(b), the decrement in the RhB absorption peak intensity can be attributed to the cleavage of the aromatic ring of the RhB dye during photocatalysis. Another intriguing feature to be noticed from these figures is the shift of the absorption peak to shorter wavelengths during photocatalysis in the presence of CsSnCl\textsubscript{3} nanocrystals. Such hypsochromic shifts can be imputed to the formation of a series of N-deethylate intermediates of RhB during photocatalysis. The trend of this blue-shift under visible and UV-visible irradiation is demonstrated in Fig. \ref{fig:photocatalytic}(d). As shown in this figure, the maximum absorption band of RhB solution significantly shifted from 553 nm to 498 nm under 240 minutes of UV-visible irradiation, whereas the peak wavelength shift was relatively insignificant under visible irradiation. Hence, it can be inferred that the effect of N-deethylation process on photocatalysis was more prominent under UV-visible light as compared to the visible light illumination. 

\subsection{Theoretical Analysis}

To explain the underlying mechanism behind the photocatalytic activities of as-synthesized CsSnCl\textsubscript{3} nanocrystals, a comprehensive investigation is further required on their optical and electronic properties. However, the complexity of the experimental conditions and to some extent, the unavailability of the required experimental set-up pose difficulty to carry out such measurements precisely. Therefore, we have analyzed both the optical and electronic properties i.e. band structure, effective mass, density of states, atomic bond formation and electron charge density of as-synthesized CsSnCl\textsubscript{3} nanocrystals via density functional theory (DFT) based first-principles calculation by employing the experimentally obtained crystallographic parameters. \cite{Das2021}

\subsubsection{Computational details}

The DFT based first-principles calculation was carried out within the plane wave pseudopotential (PWPP) framework as implemented in the Cambridge Serial Total Energy Package (CASTEP). \cite{segall2002first} We have adopted both the standard generalized gradient approximation (GGA) and GGA+U methods based on the PBE gradient corrected exchange-correlation functional. \cite{perdew1996generalized} The crystallographic structural parameters obtained from the Rietveld refined powder XRD spectrum of as-synthesized CsSnCl\textsubscript{3} were employed for DFT calculation. The Cs 5s$^2$5p$^6$6s$^1$, Sn 5s$^2$ 5p$^2$ and Cl 3s$^2$3p$^5$ electrons were treated as the valance electrons. To relax the configuration in terms of atomic locations and lattice parameters, the Broyden-Fletcher-Goldfarb-Shannon (BFGS) algorithm was implemented. \cite{fischer1992general} The atomic structures were completely relaxed before the residual forces fell below 0.01 eV/\AA $\;$ with a self-consistent field (SCF) tolerance of 5.0$\times$10$^{-7}$ eV/atom.

Prior to calculation, convergence study and geometry optimization were carried out to select the optimum plane-wave cutoff energy and k-points. Fig. S4(a) in the supplementary information shows the plane-wave cut-off energy convergence result for structural optimization where the dashed line indicates the default cut-off energy. Noticeably, 600 eV cut-off energy was found to be sufficient to converge the ground state energy. Therefore, for the subsequent calculations, we have applied the Vanderbilt type ultrasoft pseudopotential to the plane wave basis sets with the optimized cut-off energy of 600 eV. Moreover, from Fig. S4(b), it can be observed that a 15 x 15 x 15 Monkhorst-Pack k-point mesh is sufficient to obtain the ground state energy of the CsSnCl\textsubscript{3} perovskite and hence, 15 x 15 x 15 k-point grid was employed for Brillouin-zone integration. \cite{monkhorst1976special} The convergence thresholds for the geometry optimization were: (a) total energy 5.0$\times$10$^{-6}$ eV/atom (b) maximum force 0.01 eV/\AA (c) maximum stress 0.02 GPa (d) maximum displacement 5.0$\times$10$^{-4}$ \AA. 

Finally, to investigate the effect of on-site Coulomb and exchange interactions on the structural, optical and electronic properties of CsSnCl\textsubscript{3} perovskite, the GGA+U calculation was conducted with different U\textsubscript{eff} values. To be specific, U\textsubscript{eff} was varied from 1 to 7 eV for the s and p orbitals of all the atoms of CsSnCl\textsubscript{3}. 

\subsubsection{First-principles calculation of crystal structure}

The structural parameters of CsSnCl\textsubscript{3} perovskite have been determined via first-principles calculation using GGA (U\textsubscript{eff} = 0 eV) and GGA+U (U\textsubscript{eff} = 1 to 7 eV) approaches. The calculated lattice constant $a=b=c$ along with the unit cell volume for U\textsubscript{eff} = 1 to 7 eV are presented in Table 2. For comparison, the lattice parameters and unit cell volume obtained from Rietveld refined powder XRD spectrum of CsSnCl\textsubscript{3} are also included in Table 2. As can be observed, the calculated structural parameters were found to be slightly larger than the experimentally obtained ones at all values of U\textsubscript{eff} except U\textsubscript{eff} = 7 eV for which the lattice parameter and unit cell volume deviated largely from the experimental values. Moreover, within GGA+U approximation, the calculated lattice parameters and unit cell volume increased with the increment of U\textsubscript{eff} from 1 to 7 eV. Notably, such a variation of structural parameters are consistent with previous investigations of analogous materials. \cite{Das2021, shenton2017effects}

\begin{table*}
    \centering 
    \caption{Lattice parameters and unit cell volume of CsSnCl\textsubscript{3} for different values of U\textsubscript{eff} obtained through first-principles calculation along with the corresponding experimental values.}
    \begin{tabular}{c c c c c c c c c c}
    \hline
         &  Experimental & U\textsubscript{eff} = & U\textsubscript{eff} = & U\textsubscript{eff} = & U\textsubscript{eff} = & U\textsubscript{eff} = & U\textsubscript{eff} = & U\textsubscript{eff} = & U\textsubscript{eff} = \\
         & value & 0 eV & 1 eV & 2 eV & 3 eV & 4 eV& 5 eV & 6 eV & 7 eV\\
    \hline
    $a=b=c$ (\AA) & 5.583 & 5.613 & 5.605 & 5.605 & 5.605 & 5.625 & 5.739 & 5.890 & 6.348 \\
    Volume (\AA$^3$) & 174.02 & 176.84 & 175.89 & 175.89 & 175.89 & 177.97 & 189.02 & 204.33 & 255.80 \\ 
    \hline     
         & 
    \end{tabular}
    \label{tab:ueff_vs_struct}
\end{table*}

\begin{figure*}[t]
   \centering
   \includegraphics[width=15 cm]{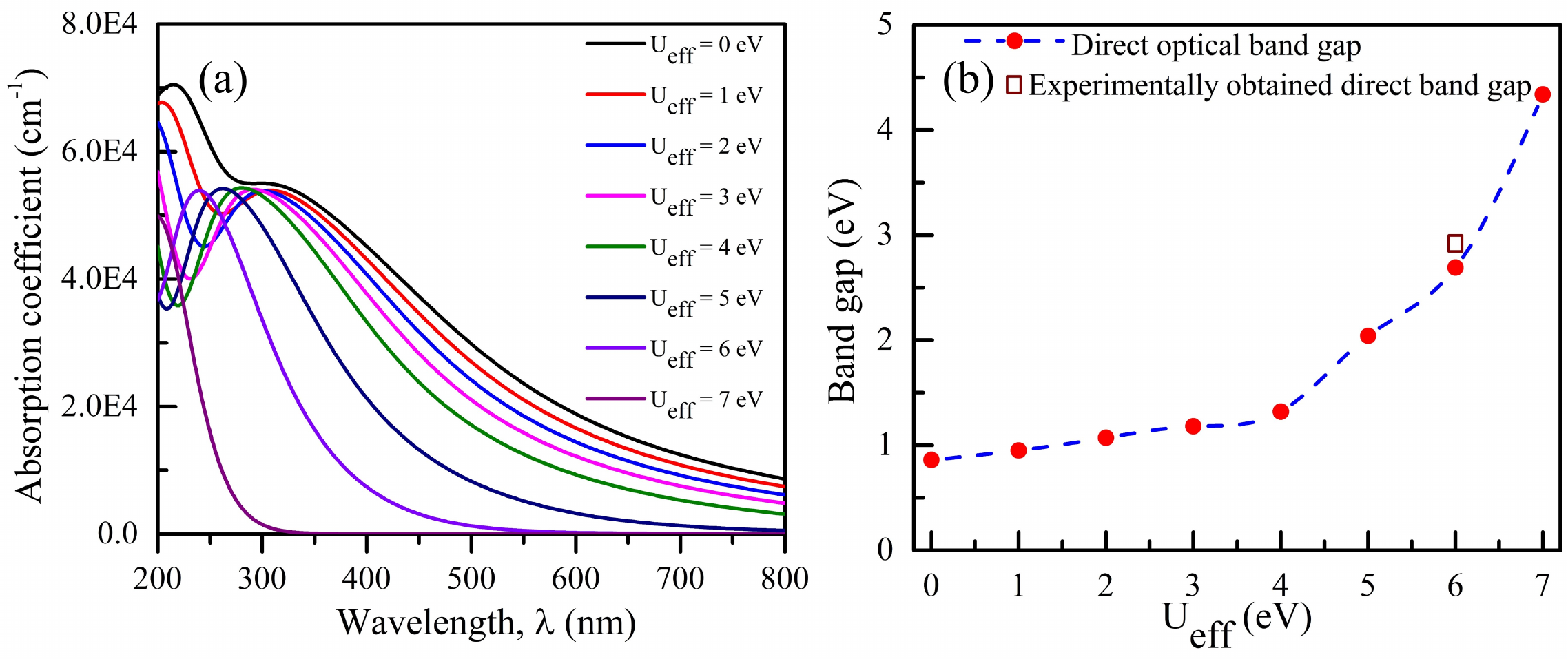}% Here is how to import EPS art
   \caption{\label{fig:optictheory} (a) Variation of theoretical absorption coefficient as a function of wavelength for different U\textsubscript{eff}. (b) Direct optical band gap as a function U\textsubscript{eff}. For U\textsubscript{eff} = 6 eV the theoretically calculated direct optical band gap matches well with the experimentally obtained direct band gap of $\sim$2.98 eV (square marked).}
\end{figure*}

\subsubsection{Theoretical analysis of optical properties}

We have calculated the optical absorption coefficient of CsSnCl\textsubscript{3} perovskite by first-principles calculation via GGA (U\textsubscript{eff} = 0 eV) and GGA+U (U\textsubscript{eff} = 1 to 7 eV) methods using the following equation:
\begin{equation}
  \alpha(\omega) = \sqrt[]{2}\omega~\sqrt[]{- \varepsilon _{1} \left(\omega  \right)~\sqrt[]{ \varepsilon _{1}^{2} \left(  \omega  \right) + \varepsilon _{2}^{2} \left(\omega\right)}} 
\end{equation}
where $\alpha(\omega)$ is the absorption coefficient, \(  \varepsilon _{1} \left(  \omega  \right)  \)  and  \(  \varepsilon _{2} \left(  \omega  \right)  \)  are the frequency dependent real and imaginary parts of the dielectric function,  \(  \omega  \)  is photon frequency. \cite{basith2018low, deng2019theoretical} The real part of the dielectric function  \(  \varepsilon _{1} \left(  \omega  \right)  \)  was evaluated from the imaginary part  \(  \varepsilon _{2} \left(  \omega  \right)  \)  by the Kramers-Kronig relationship. \cite{peng2016electronic}

Fig. \ref{fig:optictheory}(a) demonstrates the calculated absorption coefficient of CsSnCl\textsubscript{3} as a function of wavelength for different values of Hubbard U\textsubscript{eff} parameter. It can be noticed that for all values of U\textsubscript{eff}, the absorption peak of CsSnCl\textsubscript{3} lies in the UV region which is consistent with the absorbance spectrum obtained by UV-visible spectroscopy [Fig. \ref{fig:optic}(a)]. Clearly, with the increase of U\textsubscript{eff}, the theoretically calculated absorption coefficient spectra of CsSnCl\textsubscript{3} underwent a blue shift. Moreover, it is worth noticing that the trend of coefficient spectrum calculated using U\textsubscript{eff} = 7 eV is reasonably different than the other spectra for U\textsubscript{eff} = 0 to 6 eV. Such observation suggests that the U\textsubscript{eff} value should be kept below 7 eV for the GGA+U approximation to determine the optical properties and electronic band structure of CsSnCl\textsubscript{3} nanocrystals.  

Further, we have theoretically calculated the optical band gap of CsSnCl\textsubscript{3} nanocrystals for different values of U\textsubscript{eff} by employing the calculated absorption coefficient spectra in Tauc relation. The calculated direct band gap values are presented in Fig. \ref{fig:optictheory}(b) as a function of U\textsubscript{eff}. It is worth noting that upon varying the U\textsubscript{eff} values from 0 to 4 eV, we observed nominal change in the calculated band gap. Moreover, for this range of U\textsubscript{eff} values i.e. for U\textsubscript{eff} = 0 to 4 eV, the calculated direct band gap remained within 1.1 eV which is significantly smaller as compared to the band gap obtained experimentally. Interestingly, the calculated band gap value increased significantly when we applied U\textsubscript{eff}$>$ 4 eV. As can be seen in Fig. \ref{fig:optictheory}(b), the direct optical band gap value of CsSnCl\textsubscript{3} was calculated to be $\sim$2.69 eV for U\textsubscript{eff} = 6 eV which matches well with the experimental one $\sim$2.98 eV (marked by a square in the figure). This confirms the high i.e. 90\% accuracy of band gap estimation by considering GGA+U method in first-principles calculation. For a further increase of U\textsubscript{eff} to 7 eV, the optical band gap increased abruptly to $\sim$4.4 eV which is much higher than the experimental value. This observation implies that among all the U\textsubscript{eff} values employed for GGA+U approximation, U\textsubscript{eff} = 6 eV most accurately localized the orbitals of CsSnCl\textsubscript{3} nanocrystals.

\subsubsection{Electronic band structure}

Initially, with a view to bench-marking, we have determined the electronic band structure of CsSnCl\textsubscript{3} perovskite using the hybrid exchange-correlation i.e. HSE06 functional which has been illustrated in Fig. S5 of ESI. As shown in the figure, using this functional, the electronic band structure of CsSnCl\textsubscript{3} was determined to be direct with a band gap value of 1.534 eV. Notably, this outcome agrees quite well with the previously reported HSE06 functional based electronic band structure of CsSnCl\textsubscript{3} \cite{korbel2016stability} and thus, validate the simulation framework of the present investigation. However, this band gap value (1.534 eV) is almost 45.9\% smaller than the experimentally obtained optical band gap of CsSnCl\textsubscript{3} nanocrystals. Notably, according to a previous investigation, \cite{bredas2014mind} the electronic band gap of any material would be comparable with its optical band gap. Such an underestimation of the band gap of CsSnCl\textsubscript{3} nanocrystals reveals the severe limitation of HSE06 functional based first-principles calculation to accurately analyze the electronic band structure of related all-inorganic halide perovskites. 

Therefore, further, we have analyzed the band structure of CsSnCl\textsubscript{3} perovskite using the GGA (U\textsubscript{eff} = 0 eV) method incorporating PBE functional as demonstrated in ESI Fig. S6(a). The dashed line between the valence and conduction bands indicates the Fermi level. As can be seen, we obtained a direct electronic band gap of 0.95 eV which is much smaller than the direct optical band gap $\sim$2.98 eV of fabricated CsSnCl\textsubscript{3} nanocrystals as obtained by UV-visible and PL spectroscopic analyses. Hence, we may infer that the band structure formation of CsSnCl\textsubscript{3} was erroneous for U\textsubscript{eff} = 0 eV.  Therefore, we have calculated the electronic band structure via GGA+U method with U\textsubscript{eff} = 1 to 7 eV and presented the result in ESI Fig. S6(b) to S6(h). It can be clearly seen in Fig. S6 that for all values of U\textsubscript{eff}, both the VBM and CBM lie at the same symmetry point R confirming the direct band structure of CsSnCl\textsubscript{3}. However, the band gap enlarged with increasing U\textsubscript{eff} which can be associated with the enhanced localization of the Cl 3p and Sn 5p orbitals owing to increased U\textsubscript{eff}. It is also worth mentioning that the calculated electronic band structure was semiconducting while the U\textsubscript{eff} value remained within 6 eV. However, for U\textsubscript{eff} = 7 eV, the band structure became insulating which is contradictory to the experimental result. Therefore, we may recommend an U\textsubscript{eff} value of 6 eV for the first-principles calculation of CsSnCl\textsubscript{3} nanocrystals.

\subsubsection{Effective mass}

It is well known that along with optical characteristics and electronic band structure, the photocatalytic performance of a typical semiconductor photocatalyst is also significantly influenced by the mobility of its photogenerated electrons and holes. Hence, to get insight into the carrier transport property of CsSnCl\textsubscript{3} nanocrystals, we have calculated its charge carrier effective masses ($m^{*}$) using the following equation- \cite{deng2019theoretical, yu2018surface}

\begin{equation}
m^{*}=\hbar^{2} \left ( \frac{\mathrm{d}^{2}E}{\mathrm{d} k^{2}} \right )^{-1}
\end{equation}

\begin{figure}[t]
   \centering
   \includegraphics[width=7.5 cm]{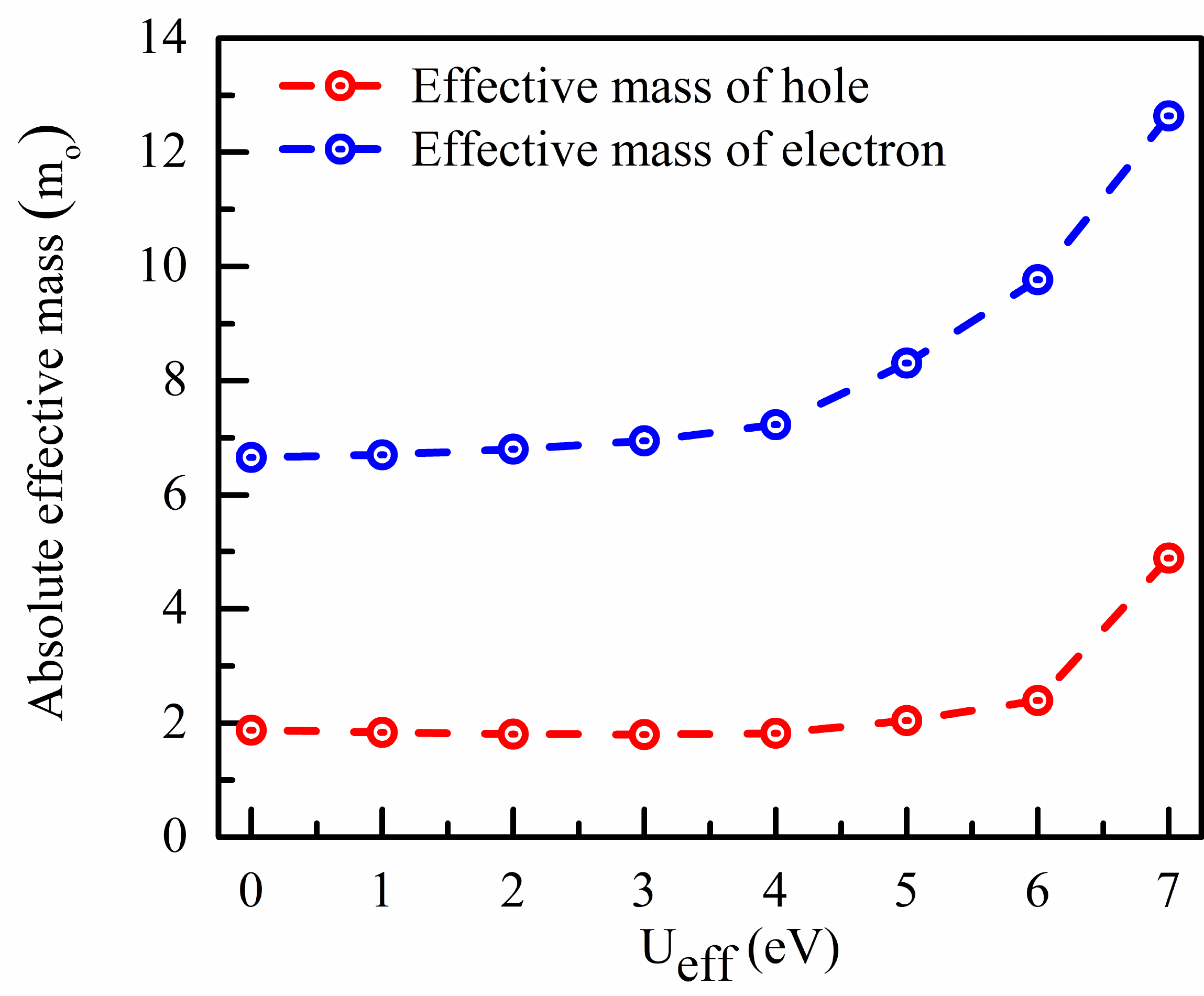}% Here is how to import EPS art
   \caption{\label{fig:effective_mass} Variation of the absolute effective masses of holes and electrons in terms of the electron rest mass, m\textsubscript{o} with the increase of U\textsubscript{eff} values.} 
\end{figure}

Here, E is the band-edge energy as a function of wave-vector k and $\hbar$ is the reduced Planck constant. The effective mass of electrons (m$_{e}$$^{*}$) and holes (m$_{h}$$^{*}$) were evaluated by parabolic fitting of the E-k curve within the small region of wave-vector near the CBM and VBM, respectively. Fig. \ref{fig:effective_mass} illustrates the variation of m$_{e}$$^{*}$ and m$_{h}$$^{*}$ of CsSnCl\textsubscript{3} nanocrystal as a function of U\textsubscript{eff}. It can be observed that m$_{h}$$^{*}$ did not change significantly with the increase of U\textsubscript{eff} up to 6 eV.  However, a steeper increase can be noticed in m$_{h}$$^{*}$ from 2.39 m\textsubscript{o} to 4.89 m\textsubscript{o} for U\textsubscript{eff} = 6 eV to U\textsubscript{eff} = 7 eV which suggests a decrease in curvature at the VBM. Similarly, in Fig. \ref{fig:effective_mass}, we can observe that the change in m$_{e}$$^{*}$ was not notable with the increase of U\textsubscript{eff} up to 6 eV. However, the calculated m$_{e}$$^{*}$ value significantly increased from 9.77 m\textsubscript{o} to 12.64 m\textsubscript{o} when we applied U\textsubscript{eff} = 7 eV which can be attributed to the decrease in the curvature of the CBM. Such variation trend again justifies that it would be rational to consider U\textsubscript{eff} below 7 eV for CsSnCl\textsubscript{3} perovskite .

Typically, lower effective mass of a semiconductor results in higher transportability of photogenerated charge carriers from its bulk to surface which can mitigate the recombination of e$^{-}$-h$^{+}$ pairs and thus, enhance photocatalytic activity. In order to get a deep insight into the recombination rate within photocatalysts, a number of previous investigations \cite{zhang2012towards, zhang2015illustration} have defined the ratio of effective mass of the electron and
hole using the following formula- 

\begin{equation} \label{eq:eff_mass_ratio}
D =\frac{\mathrm m_{h}^{*}}{\mathrm m_{e}^{*}}
\end{equation}

Notably, the aforementioned investigations have reported that materials with ``D'' values much greater than 1 demonstrate excellent photocatalytic activities. \cite{zhang2012towards} Interestingly, in this present investigation, the ``D'' value of CsSnCl\textsubscript{3} perovskite can be calculated as 0.25 for U\textsubscript{eff} = 6 eV which is much smaller than 1. However, despite possessing such a small ``D'' value, our as-synthesized CsSnCl\textsubscript{3} nanocrystals have exhibited notable photocatalytic performance without the assistance of any co-catalyst or re-agents under both visible and UV-visible irradiation [Fig. \ref{fig:photocatalytic}(c)]. Based on these outcomes, it can be inferred that as the m$_{e}$$^{*}$ value is larger than m$_{h}$$^{*}$ in CsSnCl\textsubscript{3}, a ``D'' value much smaller than 1 indicates a large difference in its electron and hole mobility which has facilitated the separation of photoexcited e$^{-}$-h$^{+}$ pairs, inhibited recombination and consequently enhanced photocatalytic activity. Therefore, it might be worthwhile to propose a modification for the interpretation of Eqn. [\ref{eq:eff_mass_ratio}] as follows. A ``D'' value either much smaller or much larger than 1 is the indication of low recombination rate and high photocatalytic efficiency. Here, for clarification, it should be mentioned that since for CsSnCl\textsubscript{3}, an U\textsubscript{eff} value of 6 eV most accurately determined the optical band gap and electronic band structure, hence, we used the effective masses obtained for U\textsubscript{eff} = 6 eV to evaluate the ``D'' value.

\subsubsection{Density of states}

\begin{figure*}[t]
   \centering
   \includegraphics[width=15 cm]{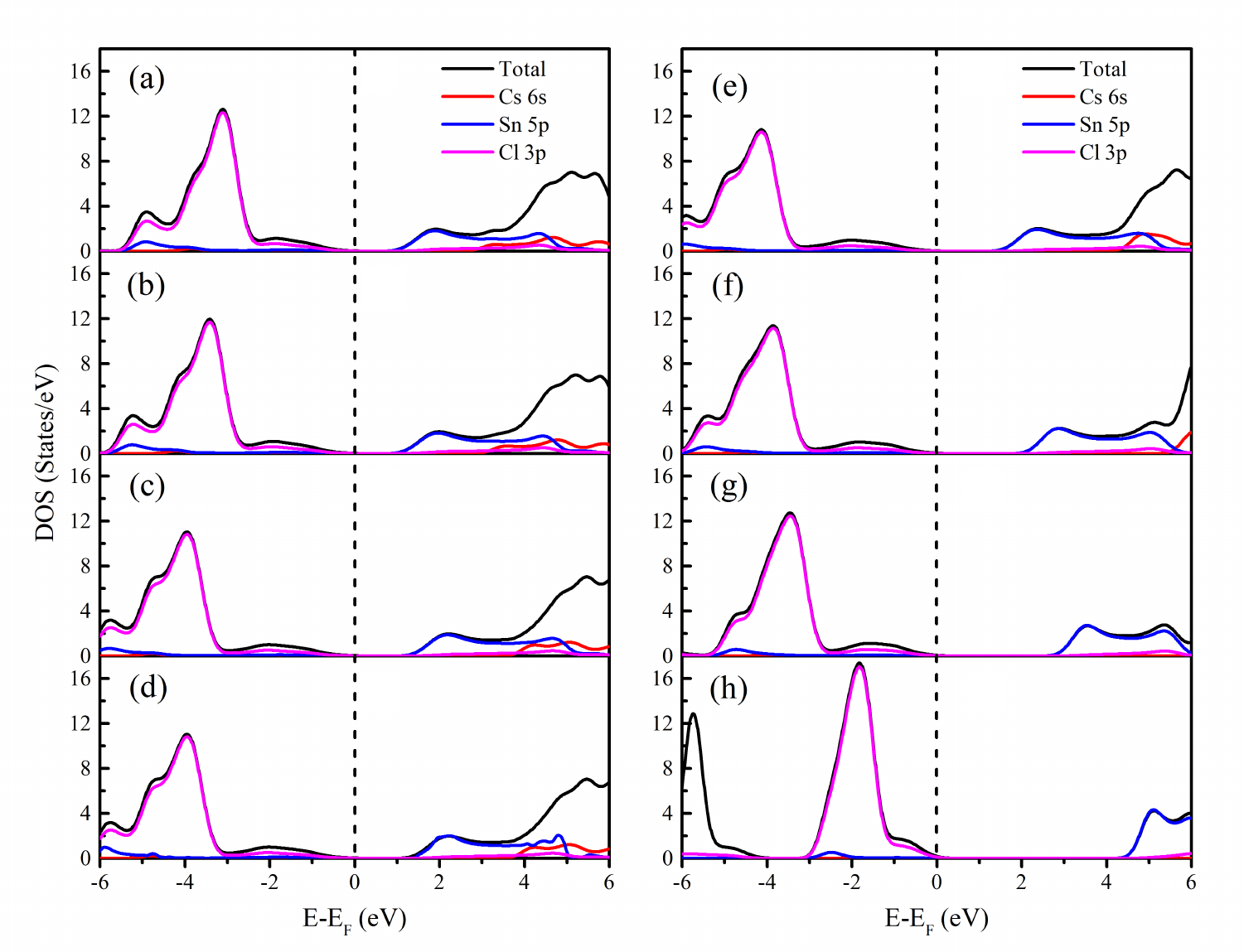}% Here is how to import EPS art
   \caption{\label{fig:dos} The calculated total density of states (TDOS) of CsSnCl\textsubscript{3} and the Cs 6s, Sn 5p and Cl 3p orbitals. The panels (a)–(h) show the DOS for U\textsubscript{eff} = 0 eV to 7 eV respectively. The zero position is set to the Fermi energy.} 
\end{figure*}
The total density of states (TDOS) and partial density of states (PDOS) for the Cs 6s, Sn 5p, and Cl 3p orbitals of CsSnCl\textsubscript{3} perovskite have been determined using the GGA (U\textsubscript{eff} = 0 eV) and GGA+U (U\textsubscript{eff} = 1 to 7 eV) methods to resolve each individual orbital’s contributions to its electronic bands. The TDOS and PDOS obtained for U\textsubscript{eff} = 0 to 7 eV are illustrated in Fig. \ref{fig:dos}(a) to (h), respectively. As can be seen in these figures, for all values of U\textsubscript{eff}, the valence band (E-E\textsubscript{F} $<$ 0 eV) of CsSnCl\textsubscript{3} is made up of the hybridization of Cl 3p and Sn 5p orbitals with  the major contribution from Cl 3p states. On the other hand, the conduction band (E-E\textsubscript{F} $>$ 0 eV) had primarily the characteristics of Sn 5p orbital with a minor contribution from Cs 6s. It should be mentioned that for U\textsubscript{eff} = 0 to 4 eV, the other orbital states of CsSnCl\textsubscript{3} had also contributed to form the TDOSs lying at the high energy region i.e. from 4 eV to 6 eV along with the Cs 6s, Sn 5p, and Cl 3p orbitals. However, to keep the figures simple and understandable, we did not demonstrate the contribution from the other orbitals. Further, as can be clearly seen in the Fig. \ref{fig:dos}(a) to (h), with the increase of U\textsubscript{eff} from 1 to 7 eV, the conduction band shifted to higher energy resulting in the enlargement of band gap. Interestingly, at U\textsubscript{eff} = 7 eV (Fig. \ref{fig:dos}(h)), we observed that the computed PDOS for Cl 3p orbital was significantly different than the PDOSs calculated for other U\textsubscript{eff} values which is another indication for the limitation of U\textsubscript{eff} $>$ 6 eV to explain the electronic band structure of CsSnCl\textsubscript{3} perovskite accurately.

\subsubsection{Electron charge density}

\begin{figure}
   \centering
   \includegraphics[width=7.5 cm]{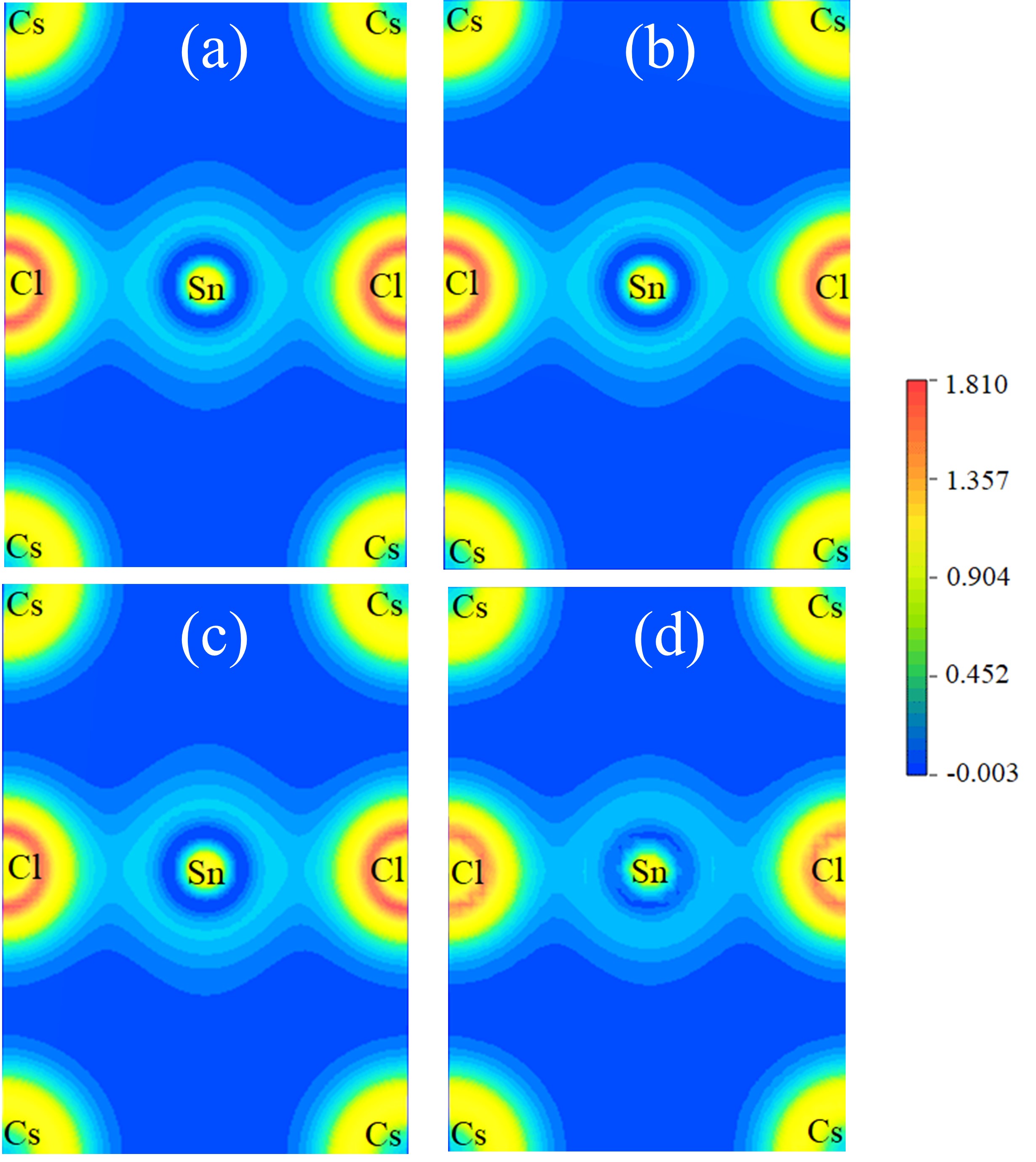}% Here is how to import EPS art
   \caption{\label{fig:chargedensity} Electronic charge density along (110) plane of CsSnCl\textsubscript{3} for (a) U\textsubscript{eff} = 0 eV, (b) U\textsubscript{eff} = 3 eV, (c) U\textsubscript{eff} = 6 eV, and (d) U\textsubscript{eff} = 7 eV.} 
\end{figure}

The electron charge density distribution of CsSnCl\textsubscript{3} perovskite may provide valuable information about its chemical bonding characteristics. Hence, we have determined the charge density map of CsSnCl\textsubscript{3} along the (110) plane using the GGA (U\textsubscript{eff} = 0 eV) and GGA+U (U\textsubscript{eff} = 3, 6 and 7 eV) methods and illustrated the result in Fig. \ref{fig:chargedensity}. It is well-known that when the electron clouds from two atoms overlap with each other and the electrons concentrate in the overlapping region, a covalent bond is formed between them. \cite{phillips1968covalent} In Fig.  \ref{fig:chargedensity}(a) to \ref{fig:chargedensity}(d), a significant overlapping of electron clouds can be noticed between Sn and Cl atoms which confirms the formation of the sigma-type Sn-Cl covalent bond in the CsSnCl\textsubscript{3} perovskite. \cite{dalpian2017changes} As illustrated by the DOS plots in Fig. \ref{fig:dos}, the charge sharing between Sn-Cl can be associated with the hybridization of Sn 5p and Cl 3p orbitals. 

For a deeper insight into the chemical bonding nature, we have performed Mulliken population analysis and calculated the effective atomic charge, bond population and bond length in as-synthesized CsSnCl\textsubscript{3} crystal structure. The results have been explained in details in the ESI. Notably, our analysis demonstrated that the degree of bond covalency in CsSnCl\textsubscript{3} reduces with increasing U\textsubscript{eff} due to the enhanced effect of on-site Coulomb interaction.

\section{Conclusions}

We have demonstrated a rapid, facile hot-injection technique for the synthesis of lead-free non-toxic CsSnCl\textsubscript{3} perovskite nanocrystals having high structural stability over a wide temperature range and excellent crystallinity. The as-synthesized CsSnCl\textsubscript{3} perovskite exhibited photocatalytic performance without using any reagent under both visible and UV-visible irradiation owing to their direct band structure and favorable band edge positions. Further, density functional theory based analysis provided deep insight into the optical absorption property, electronic band structure, charge carrier effective masses and electron charge density of CsSnCl\textsubscript{3} nanocrystals at atomic level. The outcome of our theoretical analysis revealed the significance of considering effect of on-site Coulomb and exchange interaction energy in GGA+U approximation. Notably, when a Hubbard U\textsubscript{eff} correction parameter of 6 eV was employed, the GGA+U method incorporated with Perdew-Burke-Ernzerhof gradient functional provided results with almost 90\% accuracy which is much higher as compared to the previously reported conventional Heyd, Scuseria, and Ernzerhof (HSE06) hybrid exchange-correlation functional showing 45.9\% accuracy. Moreover, the ratio ``D'' of the theoretically calculated effective masses of the hole and electron of CsSnCl\textsubscript{3} nanocrystals shed light into the rationale behind their satisfactory photocatalytic performance which led us to propose a new interpretation of ``D'' value. We have suggested that low photogenerated charge-carrier recombination rate and hence, high photocatalytic activity can be predicted for materials with ``D'' value much larger or smaller than 1. The outcome of this investigation might be useful for the theoretical analysis of related halide perovskites in which the on-site Coulomb interaction energy is expected to play a vital role. Moreover, the demonstrated rapid hot-injection technique might be effectively employed to synthesized analogous all-inorganic halide perovskites with excellent structural stability and crystallinity.

\section*{Acknowledgments}
The financial assistance from the Committee for Advanced 
Studies and Research, Bangladesh
University of Engineering and Technology (BUET) is acknowledged. The computational facility provided by the IICT, BUET is also acknowledged. Sincere gratefulness to Professor Tadahiro Komeda and Dr. Ferdous Ara, Institute of Multidisciplinary Research for Advanced Materials (IMRAM), Tohoku University, Japan for XPS measurements and TEM imaging.\\   \\

\section*{Data availability}
The raw and processed data required to reproduce these findings cannot be shared at this time due to technical or time limitations.

\section*{Supplementary Information}
Additional electronic supplementary information (ESI) is available. See DOI: \href{https://doi.org/10.1039/D1CP02666F}{10.1039/D1CP02666F}

%% file: main.bbl
\providecommand*{\mcitethebibliography}{\thebibliography}
\csname @ifundefined\endcsname{endmcitethebibliography}
{\let\endmcitethebibliography\endthebibliography}{}
\begin{mcitethebibliography}{83}
\providecommand*{\natexlab}[1]{#1}
\providecommand*{\mciteSetBstSublistMode}[1]{}
\providecommand*{\mciteSetBstMaxWidthForm}[2]{}
\providecommand*{\mciteBstWouldAddEndPuncttrue}
  {\def\EndOfBibitem{\unskip.}}
\providecommand*{\mciteBstWouldAddEndPunctfalse}
  {\let\EndOfBibitem\relax}
\providecommand*{\mciteSetBstMidEndSepPunct}[3]{}
\providecommand*{\mciteSetBstSublistLabelBeginEnd}[3]{}
\providecommand*{\EndOfBibitem}{}
\mciteSetBstSublistMode{f}
\mciteSetBstMaxWidthForm{subitem}
{(\emph{\alph{mcitesubitemcount}})}
\mciteSetBstSublistLabelBeginEnd{\mcitemaxwidthsubitemform\space}
{\relax}{\relax}

\bibitem[Mao \emph{et~al.}(2018)Mao, Stoumpos, and Kanatzidis]{mao2018two}
L.~Mao, C.~C. Stoumpos and M.~G. Kanatzidis, \emph{J. Am. Chem. Soc.}, 2018,
  \textbf{141}, 1171--1190\relax
\mciteBstWouldAddEndPuncttrue
\mciteSetBstMidEndSepPunct{\mcitedefaultmidpunct}
{\mcitedefaultendpunct}{\mcitedefaultseppunct}\relax
\EndOfBibitem
\bibitem[Lin \emph{et~al.}(2017)Lin, Pattanasattayavong, and
  Anthopoulos]{lin2017metal}
Y.-H. Lin, P.~Pattanasattayavong and T.~D. Anthopoulos, \emph{Adv. Mater.},
  2017, \textbf{29}, 1702838\relax
\mciteBstWouldAddEndPuncttrue
\mciteSetBstMidEndSepPunct{\mcitedefaultmidpunct}
{\mcitedefaultendpunct}{\mcitedefaultseppunct}\relax
\EndOfBibitem
\bibitem[Yantara \emph{et~al.}(2015)Yantara, Bhaumik, Yan, Sabba, Dewi,
  Mathews, Boix, Demir, and Mhaisalkar]{yantara2015inorganic}
N.~Yantara, S.~Bhaumik, F.~Yan, D.~Sabba, H.~A. Dewi, N.~Mathews, P.~P. Boix,
  H.~V. Demir and S.~Mhaisalkar, \emph{J. Phys. Chem. Lett.}, 2015, \textbf{6},
  4360--4364\relax
\mciteBstWouldAddEndPuncttrue
\mciteSetBstMidEndSepPunct{\mcitedefaultmidpunct}
{\mcitedefaultendpunct}{\mcitedefaultseppunct}\relax
\EndOfBibitem
\bibitem[Xu \emph{et~al.}(2019)Xu, Hu, Bai, Bao, Miao, Yuan, Borzda, Barker,
  Tyukalova, Hu, Kawecki, Wang, Yan, Liu, Shi, Uvdal, Fahlman, Zhang, Duchamp,
  Liu, Petrozza, Wang, Liu, Huang, and Gao]{xu2019rational}
W.~Xu, Q.~Hu, S.~Bai, C.~Bao, Y.~Miao, Z.~Yuan, T.~Borzda, A.~J. Barker,
  E.~Tyukalova, Z.~Hu, M.~Kawecki, H.~Wang, Z.~Yan, X.~Liu, X.~Shi, K.~Uvdal,
  M.~Fahlman, W.~Zhang, M.~Duchamp, J.~M. Liu, A.~Petrozza, J.~Wang, L.~M. Liu,
  W.~Huang and F.~Gao, \emph{Nat. Photonics}, 2019, \textbf{13}, 418--424\relax
\mciteBstWouldAddEndPuncttrue
\mciteSetBstMidEndSepPunct{\mcitedefaultmidpunct}
{\mcitedefaultendpunct}{\mcitedefaultseppunct}\relax
\EndOfBibitem
\bibitem[Tian \emph{et~al.}(2016)Tian, Ling, Shu, Zhou, Besara, Siegrist, Gao,
  and Ma]{tian2016solution}
Y.~Tian, Y.~Ling, Y.~Shu, C.~Zhou, T.~Besara, T.~Siegrist, H.~Gao and B.~Ma,
  \emph{Adv. Electron. Mater.}, 2016, \textbf{2}, 1600165\relax
\mciteBstWouldAddEndPuncttrue
\mciteSetBstMidEndSepPunct{\mcitedefaultmidpunct}
{\mcitedefaultendpunct}{\mcitedefaultseppunct}\relax
\EndOfBibitem
\bibitem[Ghosh \emph{et~al.}(2021)Ghosh, Pradhan, Zhang, Hofkens, Karki, and
  Materny]{ghosh2021nature}
S.~Ghosh, B.~Pradhan, Y.~Zhang, J.~Hofkens, K.~J. Karki and A.~Materny,
  \emph{Phys. Chem. Chem. Phys.}, 2021, \textbf{23}, 3983--3992\relax
\mciteBstWouldAddEndPuncttrue
\mciteSetBstMidEndSepPunct{\mcitedefaultmidpunct}
{\mcitedefaultendpunct}{\mcitedefaultseppunct}\relax
\EndOfBibitem
\bibitem[Pandey \emph{et~al.}(2019)Pandey, Vats, Yun, Bowen, Ho-Baillie,
  Seidel, Butler, and Seok]{pandey2019mutual}
R.~Pandey, G.~Vats, J.~Yun, C.~R. Bowen, A.~W.~Y. Ho-Baillie, J.~Seidel, K.~T.
  Butler and S.~I. Seok, \emph{Adv. Mater.}, 2019, \textbf{31}, 1807376\relax
\mciteBstWouldAddEndPuncttrue
\mciteSetBstMidEndSepPunct{\mcitedefaultmidpunct}
{\mcitedefaultendpunct}{\mcitedefaultseppunct}\relax
\EndOfBibitem
\bibitem[Jung \emph{et~al.}(2017)Jung, Rhim, and Moon]{jung2017tio2}
M.-H. Jung, S.~H. Rhim and D.~Moon, \emph{Sol. Energy Mater. Sol. Cells}, 2017,
  \textbf{172}, 44--54\relax
\mciteBstWouldAddEndPuncttrue
\mciteSetBstMidEndSepPunct{\mcitedefaultmidpunct}
{\mcitedefaultendpunct}{\mcitedefaultseppunct}\relax
\EndOfBibitem
\bibitem[Wang \emph{et~al.}(2021)Wang, Gao, Zhang, Chen, Junkang, Shen, Au, Li,
  Cai, and Yin]{wang2021thickness}
B.-H. Wang, B.~Gao, J.-R. Zhang, L.~Chen, G.~Junkang, S.~Shen, C.-T. Au, K.~Li,
  M.-Q. Cai and S.-F. Yin, \emph{Phys. Chem. Chem. Phys.}, 2021, \textbf{23},
  12439--12448\relax
\mciteBstWouldAddEndPuncttrue
\mciteSetBstMidEndSepPunct{\mcitedefaultmidpunct}
{\mcitedefaultendpunct}{\mcitedefaultseppunct}\relax
\EndOfBibitem
\bibitem[Haris \emph{et~al.}(2021)Haris, Kazim, Pegu, Deepa, and
  Ahmad]{haris2021substance}
M.~P.~U. Haris, S.~Kazim, M.~Pegu, M.~Deepa and S.~Ahmad, \emph{Phys. Chem.
  Chem. Phys.}, 2021, \textbf{23}, 9049--9060\relax
\mciteBstWouldAddEndPuncttrue
\mciteSetBstMidEndSepPunct{\mcitedefaultmidpunct}
{\mcitedefaultendpunct}{\mcitedefaultseppunct}\relax
\EndOfBibitem
\bibitem[Saleem \emph{et~al.}(2021)Saleem, Yang, Batool, Sulaman, Veeramalai,
  Jiang, Tang, Cui, Tang, and Zou]{saleem2021cspbi3}
M.~I. Saleem, S.~Yang, A.~Batool, M.~Sulaman, C.~P. Veeramalai, Y.~Jiang,
  Y.~Tang, Y.~Cui, L.~Tang and B.~Zou, \emph{J. Mater. Sci. Technol.}, 2021,
  \textbf{75}, 196--204\relax
\mciteBstWouldAddEndPuncttrue
\mciteSetBstMidEndSepPunct{\mcitedefaultmidpunct}
{\mcitedefaultendpunct}{\mcitedefaultseppunct}\relax
\EndOfBibitem
\bibitem[Wu \emph{et~al.}(2019)Wu, Zhou, Meng, Xue, Zhou, Tang, and
  Zhang]{wu2019air}
G.~Wu, J.~Zhou, R.~Meng, B.~Xue, H.~Zhou, Z.~Tang and Y.~Zhang, \emph{Phys.
  Chem. Chem. Phys.}, 2019, \textbf{21}, 3106--3113\relax
\mciteBstWouldAddEndPuncttrue
\mciteSetBstMidEndSepPunct{\mcitedefaultmidpunct}
{\mcitedefaultendpunct}{\mcitedefaultseppunct}\relax
\EndOfBibitem
\bibitem[Teng \emph{et~al.}(2021)Teng, Shi, Liao, and Zhao]{teng2021first}
Q.~Teng, T.~Shi, C.~Liao and Y.-J. Zhao, \emph{J. Mater. Chem. C}, 2021,
  \textbf{9}, 982--990\relax
\mciteBstWouldAddEndPuncttrue
\mciteSetBstMidEndSepPunct{\mcitedefaultmidpunct}
{\mcitedefaultendpunct}{\mcitedefaultseppunct}\relax
\EndOfBibitem
\bibitem[Sun \emph{et~al.}(2015)Sun, Agiorgousis, Zhang, and
  Zhang]{sun2015chalcogenide}
Y.-Y. Sun, M.~L. Agiorgousis, P.~Zhang and S.~Zhang, \emph{Nano Lett.}, 2015,
  \textbf{15}, 581--585\relax
\mciteBstWouldAddEndPuncttrue
\mciteSetBstMidEndSepPunct{\mcitedefaultmidpunct}
{\mcitedefaultendpunct}{\mcitedefaultseppunct}\relax
\EndOfBibitem
\bibitem[Guo \emph{et~al.}(2021)Guo, Yuan, Zhu, Yu, Wang, Lin, Wang, Qin,
  Zhang, and Ai]{guo2021influence}
Y.~Guo, S.~Yuan, D.~Zhu, M.~Yu, H.-Y. Wang, J.~Lin, Y.~Wang, Y.~Qin, J.-P.
  Zhang and X.-C. Ai, \emph{Phys. Chem. Chem. Phys.}, 2021, \textbf{23},
  6162--6170\relax
\mciteBstWouldAddEndPuncttrue
\mciteSetBstMidEndSepPunct{\mcitedefaultmidpunct}
{\mcitedefaultendpunct}{\mcitedefaultseppunct}\relax
\EndOfBibitem
\bibitem[Jung \emph{et~al.}(2017)Jung, Lee, Walsh, and Soon]{jung2017influence}
Y.-K. Jung, J.-H. Lee, A.~Walsh and A.~Soon, \emph{Chem. Mater.}, 2017,
  \textbf{29}, 3181--3188\relax
\mciteBstWouldAddEndPuncttrue
\mciteSetBstMidEndSepPunct{\mcitedefaultmidpunct}
{\mcitedefaultendpunct}{\mcitedefaultseppunct}\relax
\EndOfBibitem
\bibitem[De~Roo \emph{et~al.}(2016)De~Roo, Ib{\'a}{\~n}ez, Geiregat, Nedelcu,
  Walravens, Maes, Martins, Van~Driessche, Kovalenko, and Hens]{de2016highly}
J.~De~Roo, M.~Ib{\'a}{\~n}ez, P.~Geiregat, G.~Nedelcu, W.~Walravens, J.~Maes,
  J.~C. Martins, I.~Van~Driessche, M.~V. Kovalenko and Z.~Hens, \emph{ACS
  Nano}, 2016, \textbf{10}, 2071--2081\relax
\mciteBstWouldAddEndPuncttrue
\mciteSetBstMidEndSepPunct{\mcitedefaultmidpunct}
{\mcitedefaultendpunct}{\mcitedefaultseppunct}\relax
\EndOfBibitem
\bibitem[Protesescu \emph{et~al.}(2015)Protesescu, Yakunin, Bodnarchuk, Krieg,
  Caputo, Hendon, Yang, Walsh, and Kovalenko]{protesescu2015nanocrystals}
L.~Protesescu, S.~Yakunin, M.~I. Bodnarchuk, F.~Krieg, R.~Caputo, C.~H. Hendon,
  R.~X. Yang, A.~Walsh and M.~V. Kovalenko, \emph{Nano Lett.}, 2015,
  \textbf{15}, 3692--3696\relax
\mciteBstWouldAddEndPuncttrue
\mciteSetBstMidEndSepPunct{\mcitedefaultmidpunct}
{\mcitedefaultendpunct}{\mcitedefaultseppunct}\relax
\EndOfBibitem
\bibitem[Liang \emph{et~al.}(2018)Liang, Liu, Qiu, Hawash, Meng, Wu, Jiang,
  Ono, and Qi]{liang2018enhancing}
J.~Liang, Z.~Liu, L.~Qiu, Z.~Hawash, L.~Meng, Z.~Wu, Y.~Jiang, L.~K. Ono and
  Y.~Qi, \emph{Adv. Energy Mater.}, 2018, \textbf{8}, 1800504\relax
\mciteBstWouldAddEndPuncttrue
\mciteSetBstMidEndSepPunct{\mcitedefaultmidpunct}
{\mcitedefaultendpunct}{\mcitedefaultseppunct}\relax
\EndOfBibitem
\bibitem[Kang and Han(2018)]{kang2018intrinsic}
Y.~Kang and S.~Han, \emph{Phys. Rev. Appl.}, 2018, \textbf{10}, 044013\relax
\mciteBstWouldAddEndPuncttrue
\mciteSetBstMidEndSepPunct{\mcitedefaultmidpunct}
{\mcitedefaultendpunct}{\mcitedefaultseppunct}\relax
\EndOfBibitem
\bibitem[J.~Jia and Yang(2021)]{jia2021unfused}
L.~Z. J.~Jia, F.~Wu and C.~Yang, \emph{ACS Appl. Mater. Interfaces}, 2021,
  \textbf{13}, 33328--33334\relax
\mciteBstWouldAddEndPuncttrue
\mciteSetBstMidEndSepPunct{\mcitedefaultmidpunct}
{\mcitedefaultendpunct}{\mcitedefaultseppunct}\relax
\EndOfBibitem
\bibitem[Han \emph{et~al.}(2019)Han, Sun, Peng, Han, Chen, Liu, Guo, Zhao,
  Shan, Xu, Hao, Hu, and Yang]{han2019controllable}
M.~Han, J.~Sun, M.~Peng, N.~Han, Z.~Chen, D.~Liu, Y.~Guo, S.~Zhao, C.~Shan,
  T.~Xu, X.~Hao, W.~Hu and Z.~Yang, \emph{J. Phys. Chem. C}, 2019,
  \textbf{123}, 17566--17573\relax
\mciteBstWouldAddEndPuncttrue
\mciteSetBstMidEndSepPunct{\mcitedefaultmidpunct}
{\mcitedefaultendpunct}{\mcitedefaultseppunct}\relax
\EndOfBibitem
\bibitem[Mu \emph{et~al.}(2020)Mu, Hu, Wang, Jia, and Xiao]{mu2020effects}
H.~Mu, F.~Hu, R.~Wang, J.~Jia and S.~Xiao, \emph{J. Lumin.}, 2020,
  \textbf{226}, 117493\relax
\mciteBstWouldAddEndPuncttrue
\mciteSetBstMidEndSepPunct{\mcitedefaultmidpunct}
{\mcitedefaultendpunct}{\mcitedefaultseppunct}\relax
\EndOfBibitem
\bibitem[Xing \emph{et~al.}(2016)Xing, Kumar, Chong, Liu, Cai, Ding, Asta,
  Gr{\"a}tzel, Mhaisalkar, Mathews, and Sum]{xing2016solution}
G.~Xing, M.~H. Kumar, W.~K. Chong, X.~Liu, Y.~Cai, H.~Ding, M.~Asta,
  M.~Gr{\"a}tzel, S.~Mhaisalkar, N.~Mathews and T.~C. Sum, \emph{Adv. Mater.},
  2016, \textbf{28}, 8191--8196\relax
\mciteBstWouldAddEndPuncttrue
\mciteSetBstMidEndSepPunct{\mcitedefaultmidpunct}
{\mcitedefaultendpunct}{\mcitedefaultseppunct}\relax
\EndOfBibitem
\bibitem[Li \emph{et~al.}(2018)Li, Long, Xia, and Mi]{li2018all}
B.~Li, R.~Long, Y.~Xia and Q.~Mi, \emph{Angew. Chem., Int. Ed.}, 2018,
  \textbf{57}, 13154--13158\relax
\mciteBstWouldAddEndPuncttrue
\mciteSetBstMidEndSepPunct{\mcitedefaultmidpunct}
{\mcitedefaultendpunct}{\mcitedefaultseppunct}\relax
\EndOfBibitem
\bibitem[Siddik \emph{et~al.}(2021)Siddik, Haldar, Paul, Das, Barman, Roy, and
  Sarkar]{siddik2021nonvolatile}
A.~Siddik, P.~K. Haldar, T.~Paul, U.~Das, A.~Barman, A.~Roy and P.~K. Sarkar,
  \emph{Nanoscale}, 2021, \textbf{13}, 8864--8874\relax
\mciteBstWouldAddEndPuncttrue
\mciteSetBstMidEndSepPunct{\mcitedefaultmidpunct}
{\mcitedefaultendpunct}{\mcitedefaultseppunct}\relax
\EndOfBibitem
\bibitem[Zhou(2020)]{zhou2020perovskite}
Y.~Zhou, \emph{{Perovskite Quantum Dots: Synthesis, Properties and
  Applications}}, Springer Nature, {S}ingapore, 2020\relax
\mciteBstWouldAddEndPuncttrue
\mciteSetBstMidEndSepPunct{\mcitedefaultmidpunct}
{\mcitedefaultendpunct}{\mcitedefaultseppunct}\relax
\EndOfBibitem
\bibitem[Xia \emph{et~al.}(2020)Xia, Li, Huang, Ji, Yang, and
  Zhao]{xia2020room}
T.~Xia, Y.~Li, L.~Huang, W.~Ji, M.~Yang and X.~Zhao, \emph{ACS Appl. Mater.
  Interfaces}, 2020, \textbf{12}, 18634--18641\relax
\mciteBstWouldAddEndPuncttrue
\mciteSetBstMidEndSepPunct{\mcitedefaultmidpunct}
{\mcitedefaultendpunct}{\mcitedefaultseppunct}\relax
\EndOfBibitem
\bibitem[Wu \emph{et~al.}(2019)Wu, Zhang, Li, Shi, Xu, Chen, Ning, and
  Mi]{wu2019stabilizing}
Z.~Wu, Q.~Zhang, B.~Li, Z.~Shi, K.~Xu, Y.~Chen, Z.~Ning and Q.~Mi, \emph{Chem.
  Mater.}, 2019, \textbf{31}, 4999--5004\relax
\mciteBstWouldAddEndPuncttrue
\mciteSetBstMidEndSepPunct{\mcitedefaultmidpunct}
{\mcitedefaultendpunct}{\mcitedefaultseppunct}\relax
\EndOfBibitem
\bibitem[Yamada \emph{et~al.}(1998)Yamada, Kuranaga, Ueda, Goto, Okuda, and
  Furukawa]{yamada1998phase}
K.~Yamada, Y.~Kuranaga, K.~Ueda, S.~Goto, T.~Okuda and Y.~Furukawa, \emph{Bull.
  Chem. Soc. Jpn.}, 1998, \textbf{71}, 127--134\relax
\mciteBstWouldAddEndPuncttrue
\mciteSetBstMidEndSepPunct{\mcitedefaultmidpunct}
{\mcitedefaultendpunct}{\mcitedefaultseppunct}\relax
\EndOfBibitem
\bibitem[Peedikakkandy and Bhargava(2016)]{peedikakkandy2016composition}
L.~Peedikakkandy and P.~Bhargava, \emph{RSC Adv.}, 2016, \textbf{6},
  19857--19860\relax
\mciteBstWouldAddEndPuncttrue
\mciteSetBstMidEndSepPunct{\mcitedefaultmidpunct}
{\mcitedefaultendpunct}{\mcitedefaultseppunct}\relax
\EndOfBibitem
\bibitem[Chen \emph{et~al.}(2016)Chen, Lee, Chuang, Wu, and
  Chen]{chen2016synthesis}
L.-J. Chen, C.-R. Lee, Y.-J. Chuang, Z.-H. Wu and C.~Chen, \emph{J. Phys. Chem.
  Lett.}, 2016, \textbf{7}, 5028--5035\relax
\mciteBstWouldAddEndPuncttrue
\mciteSetBstMidEndSepPunct{\mcitedefaultmidpunct}
{\mcitedefaultendpunct}{\mcitedefaultseppunct}\relax
\EndOfBibitem
\bibitem[Wang \emph{et~al.}(2017)Wang, Guo, Muhammad, and
  Deng]{wang2017controlled}
A.~Wang, Y.~Guo, F.~Muhammad and Z.~Deng, \emph{Chem. Mater.}, 2017,
  \textbf{29}, 6493--6501\relax
\mciteBstWouldAddEndPuncttrue
\mciteSetBstMidEndSepPunct{\mcitedefaultmidpunct}
{\mcitedefaultendpunct}{\mcitedefaultseppunct}\relax
\EndOfBibitem
\bibitem[Kang \emph{et~al.}(2021)Kang, Rao, Fang, Zeng, Pan, and
  Zhong]{kang2021antioxidative}
C.~Kang, H.~Rao, Y.~Fang, J.~Zeng, Z.~Pan and X.~Zhong, \emph{Angew. Chem.},
  2021, \textbf{133}, 670--675\relax
\mciteBstWouldAddEndPuncttrue
\mciteSetBstMidEndSepPunct{\mcitedefaultmidpunct}
{\mcitedefaultendpunct}{\mcitedefaultseppunct}\relax
\EndOfBibitem
\bibitem[Idrissi \emph{et~al.}(2021)Idrissi, Ziti, Labrim, and
  Bahmad]{idrissi2021band}
S.~Idrissi, S.~Ziti, H.~Labrim and L.~Bahmad, \emph{Mater. Sci. Semicond.
  Process.}, 2021, \textbf{122}, 105484\relax
\mciteBstWouldAddEndPuncttrue
\mciteSetBstMidEndSepPunct{\mcitedefaultmidpunct}
{\mcitedefaultendpunct}{\mcitedefaultseppunct}\relax
\EndOfBibitem
\bibitem[K{\"o}rbel \emph{et~al.}(2016)K{\"o}rbel, Marques, and
  Botti]{korbel2016stability}
S.~K{\"o}rbel, M.~A.~L. Marques and S.~Botti, \emph{J. Mater. Chem. C}, 2016,
  \textbf{4}, 3157--3167\relax
\mciteBstWouldAddEndPuncttrue
\mciteSetBstMidEndSepPunct{\mcitedefaultmidpunct}
{\mcitedefaultendpunct}{\mcitedefaultseppunct}\relax
\EndOfBibitem
\bibitem[Heyd \emph{et~al.}(2003)Heyd, Scuseria, and Ernzerhof]{heyd2003hybrid}
J.~Heyd, G.~E. Scuseria and M.~Ernzerhof, \emph{J. Chem. Phys.}, 2003,
  \textbf{118}, 8207--8215\relax
\mciteBstWouldAddEndPuncttrue
\mciteSetBstMidEndSepPunct{\mcitedefaultmidpunct}
{\mcitedefaultendpunct}{\mcitedefaultseppunct}\relax
\EndOfBibitem
\bibitem[Huang and Lambrecht(2013)]{huang2013electronic}
L.-y. Huang and W.~R. Lambrecht, \emph{Phys. Rev. B}, 2013, \textbf{88},
  165203\relax
\mciteBstWouldAddEndPuncttrue
\mciteSetBstMidEndSepPunct{\mcitedefaultmidpunct}
{\mcitedefaultendpunct}{\mcitedefaultseppunct}\relax
\EndOfBibitem
\bibitem[Das \emph{et~al.}(2021)Das, Bhuyan, and Basith]{Das2021}
S.~Das, M.~D.~I. Bhuyan and M.~A. Basith, \emph{J. Mater. Res. Technol.}, 2021,
  \textbf{13}, 2408--2418\relax
\mciteBstWouldAddEndPuncttrue
\mciteSetBstMidEndSepPunct{\mcitedefaultmidpunct}
{\mcitedefaultendpunct}{\mcitedefaultseppunct}\relax
\EndOfBibitem
\bibitem[Lin \emph{et~al.}(2021)Lin, He, Hlevyack, Chen, Mo, Chou, and
  Chiang]{lin2021coherent}
M.-K. Lin, T.~He, J.~A. Hlevyack, P.~Chen, S.-K. Mo, M.-Y. Chou and T.-C.
  Chiang, \emph{ACS nano}, 2021, \textbf{15}, 3359--3364\relax
\mciteBstWouldAddEndPuncttrue
\mciteSetBstMidEndSepPunct{\mcitedefaultmidpunct}
{\mcitedefaultendpunct}{\mcitedefaultseppunct}\relax
\EndOfBibitem
\bibitem[Prasad \emph{et~al.}(2020)Prasad, Sadhukhan, Hansen, Felser, Kanungo,
  and Jansen]{prasad2020synthesis}
B.~E. Prasad, S.~Sadhukhan, T.~C. Hansen, C.~Felser, S.~Kanungo and M.~Jansen,
  \emph{Phys. Rev. Mater.}, 2020, \textbf{4}, 024418\relax
\mciteBstWouldAddEndPuncttrue
\mciteSetBstMidEndSepPunct{\mcitedefaultmidpunct}
{\mcitedefaultendpunct}{\mcitedefaultseppunct}\relax
\EndOfBibitem
\bibitem[Wang \emph{et~al.}(2006)Wang, Maxisch, and Ceder]{wang2006oxidation}
L.~Wang, T.~Maxisch and G.~Ceder, \emph{Phys. Rev. B}, 2006, \textbf{73},
  195107\relax
\mciteBstWouldAddEndPuncttrue
\mciteSetBstMidEndSepPunct{\mcitedefaultmidpunct}
{\mcitedefaultendpunct}{\mcitedefaultseppunct}\relax
\EndOfBibitem
\bibitem[Dudarev \emph{et~al.}(1998)Dudarev, Botton, Savrasov, Humphreys, and
  Sutton]{dudarev1998electron}
S.~L. Dudarev, G.~A. Botton, S.~Y. Savrasov, C.~J. Humphreys and A.~P. Sutton,
  \emph{Phys. Rev. B}, 1998, \textbf{57}, 1505\relax
\mciteBstWouldAddEndPuncttrue
\mciteSetBstMidEndSepPunct{\mcitedefaultmidpunct}
{\mcitedefaultendpunct}{\mcitedefaultseppunct}\relax
\EndOfBibitem
\bibitem[Murray \emph{et~al.}(1993)Murray, Norris, and
  Bawendi]{murray1993synthesis}
C.~B. Murray, D.~J. Norris and M.~G. Bawendi, \emph{J. Am. Chem. Soc.}, 1993,
  \textbf{115}, 8706--8715\relax
\mciteBstWouldAddEndPuncttrue
\mciteSetBstMidEndSepPunct{\mcitedefaultmidpunct}
{\mcitedefaultendpunct}{\mcitedefaultseppunct}\relax
\EndOfBibitem
\bibitem[Yu and Peng(2002)]{yu2002formation}
W.~W. Yu and X.~Peng, \emph{Angew. Chem., Int. Ed.}, 2002, \textbf{41},
  2368--2371\relax
\mciteBstWouldAddEndPuncttrue
\mciteSetBstMidEndSepPunct{\mcitedefaultmidpunct}
{\mcitedefaultendpunct}{\mcitedefaultseppunct}\relax
\EndOfBibitem
\bibitem[Rodriguez-Carvajal(1990)]{rodriguez1990fullprof}
J.~Rodriguez-Carvajal, FULLPROF: a program for Rietveld refinement and pattern
  matching analysis, Abstracts of the Satellite meeting on powder diffraction
  of the XV congress of the IUCr, Toulouse, France, 1990\relax
\mciteBstWouldAddEndPuncttrue
\mciteSetBstMidEndSepPunct{\mcitedefaultmidpunct}
{\mcitedefaultendpunct}{\mcitedefaultseppunct}\relax
\EndOfBibitem
\bibitem[Dastidar \emph{et~al.}(2017)Dastidar, Hawley, Dillon, Gutierrez-Perez,
  Spanier, and Fafarman]{dastidar2017quantitative}
S.~Dastidar, C.~J. Hawley, A.~D. Dillon, A.~D. Gutierrez-Perez, J.~E. Spanier
  and A.~T. Fafarman, \emph{J. Phys. Chem. Lett.}, 2017, \textbf{8},
  1278--1282\relax
\mciteBstWouldAddEndPuncttrue
\mciteSetBstMidEndSepPunct{\mcitedefaultmidpunct}
{\mcitedefaultendpunct}{\mcitedefaultseppunct}\relax
\EndOfBibitem
\bibitem[Hoffman \emph{et~al.}(2017)Hoffman, Zaiats, Wappes, and
  Kamat]{hoffman2017cspbbr3}
J.~B. Hoffman, G.~Zaiats, I.~Wappes and P.~V. Kamat, \emph{Chem. Mater.}, 2017,
  \textbf{29}, 9767--9774\relax
\mciteBstWouldAddEndPuncttrue
\mciteSetBstMidEndSepPunct{\mcitedefaultmidpunct}
{\mcitedefaultendpunct}{\mcitedefaultseppunct}\relax
\EndOfBibitem
\bibitem[Degen(1997)]{degen1997tables}
I.~Degen, \emph{{Tables of characteristic group frequencies for the
  interpretation of infrared and Raman spectra}}, Acolyte Publ, United Kingdom,
  1997\relax
\mciteBstWouldAddEndPuncttrue
\mciteSetBstMidEndSepPunct{\mcitedefaultmidpunct}
{\mcitedefaultendpunct}{\mcitedefaultseppunct}\relax
\EndOfBibitem
\bibitem[Feng \emph{et~al.}(2009)Feng, Tang, and Xiao]{feng2009anodization}
S.~Feng, Y.~Tang and T.~Xiao, \emph{J. Phys. Chem. C}, 2009, \textbf{113},
  4809--4813\relax
\mciteBstWouldAddEndPuncttrue
\mciteSetBstMidEndSepPunct{\mcitedefaultmidpunct}
{\mcitedefaultendpunct}{\mcitedefaultseppunct}\relax
\EndOfBibitem
\bibitem[Delgado-Mellado \emph{et~al.}(2018)Delgado-Mellado, Larriba, Navarro,
  Rigual, Ayuso, Garc{\'\i}a, and Rodr{\'\i}guez]{delgado2018thermal}
N.~Delgado-Mellado, M.~Larriba, P.~Navarro, V.~Rigual, M.~Ayuso, J.~Garc{\'\i}a
  and F.~Rodr{\'\i}guez, \emph{J. Mol. Liq.}, 2018, \textbf{260}, 37--43\relax
\mciteBstWouldAddEndPuncttrue
\mciteSetBstMidEndSepPunct{\mcitedefaultmidpunct}
{\mcitedefaultendpunct}{\mcitedefaultseppunct}\relax
\EndOfBibitem
\bibitem[Hills-Kimball \emph{et~al.}(2017)Hills-Kimball, Nagaoka, Cao,
  Chaykovsky, and Chen]{hills2017synthesis}
K.~Hills-Kimball, Y.~Nagaoka, C.~Cao, E.~Chaykovsky and O.~Chen, \emph{J.
  Mater. Chem. C}, 2017, \textbf{5}, 5680--5684\relax
\mciteBstWouldAddEndPuncttrue
\mciteSetBstMidEndSepPunct{\mcitedefaultmidpunct}
{\mcitedefaultendpunct}{\mcitedefaultseppunct}\relax
\EndOfBibitem
\bibitem[Kim \emph{et~al.}(2021)Kim, Cortecchia, Borzda, Mr\'oz, Leoncino,
  Dellasega, Lee, and Petrozza]{kim2021coordinating}
M.~Kim, D.~Cortecchia, T.~Borzda, W.~Mr\'oz, L.~Leoncino, D.~Dellasega, S.-H.
  Lee and A.~Petrozza, \emph{Chem. Mater.}, 2021, \textbf{33}, 547--553\relax
\mciteBstWouldAddEndPuncttrue
\mciteSetBstMidEndSepPunct{\mcitedefaultmidpunct}
{\mcitedefaultendpunct}{\mcitedefaultseppunct}\relax
\EndOfBibitem
\bibitem[Morales \emph{et~al.}(2013)Morales, Zayas, Contreras, and
  Salgado]{morales2013effect}
F.~L. Morales, T.~Zayas, O.~E. Contreras and L.~Salgado, \emph{Front. Mater.
  Sci.}, 2013, \textbf{7}, 387--395\relax
\mciteBstWouldAddEndPuncttrue
\mciteSetBstMidEndSepPunct{\mcitedefaultmidpunct}
{\mcitedefaultendpunct}{\mcitedefaultseppunct}\relax
\EndOfBibitem
\bibitem[Jellicoe \emph{et~al.}(2016)Jellicoe, Richter, Glass, Tabachnyk,
  Brady, Dutton, Rao, Friend, Credgington, Greenham, and
  L.~B{\"o}hm]{jellicoe2016synthesis}
T.~C. Jellicoe, J.~M. Richter, H.~F.~J. Glass, M.~Tabachnyk, R.~Brady, S.~E.
  Dutton, A.~Rao, R.~H. Friend, D.~Credgington, N.~C. Greenham and
  M.~L.~B{\"o}hm, \emph{J. Am. Chem. Soc.}, 2016, \textbf{138},
  2941--2944\relax
\mciteBstWouldAddEndPuncttrue
\mciteSetBstMidEndSepPunct{\mcitedefaultmidpunct}
{\mcitedefaultendpunct}{\mcitedefaultseppunct}\relax
\EndOfBibitem
\bibitem[Lin \emph{et~al.}(2019)Lin, Su, and Lin]{lin2019phase}
T.-W. Lin, C.~Su and C.~C. Lin, \emph{J. Inf. Disp.}, 2019, \textbf{20},
  209--216\relax
\mciteBstWouldAddEndPuncttrue
\mciteSetBstMidEndSepPunct{\mcitedefaultmidpunct}
{\mcitedefaultendpunct}{\mcitedefaultseppunct}\relax
\EndOfBibitem
\bibitem[Musazay(2015)]{musazay2015experimental}
A.~Musazay, \emph{PhD thesis}, University of Helsinki, 2015\relax
\mciteBstWouldAddEndPuncttrue
\mciteSetBstMidEndSepPunct{\mcitedefaultmidpunct}
{\mcitedefaultendpunct}{\mcitedefaultseppunct}\relax
\EndOfBibitem
\bibitem[Zhang \emph{et~al.}(2019)Zhang, Tai, Zhou, Zhou, Wei, Tan, Wu, Li, and
  Lin]{zhang2019enhanced}
Q.~Zhang, M.~Tai, Y.~Zhou, Y.~Zhou, Y.~Wei, C.~Tan, Z.~Wu, J.~Li and H.~Lin,
  \emph{ACS Sustainable Chem. Eng.}, 2019, \textbf{8}, 1219--1229\relax
\mciteBstWouldAddEndPuncttrue
\mciteSetBstMidEndSepPunct{\mcitedefaultmidpunct}
{\mcitedefaultendpunct}{\mcitedefaultseppunct}\relax
\EndOfBibitem
\bibitem[Allioux \emph{et~al.}(2020)Allioux, Merhebi, Ghasemian, Tang, Merenda,
  Abbasi, Mayyas, Daeneke, O’Mullane, Daiyan, Amal, and
  Kalantar-Zadeh]{allioux2020bi}
F.-M. Allioux, S.~Merhebi, M.~B. Ghasemian, J.~Tang, A.~Merenda, R.~Abbasi,
  M.~Mayyas, T.~Daeneke, A.~P. O’Mullane, R.~Daiyan, R.~Amal and
  K.~Kalantar-Zadeh, \emph{Nano Lett.}, 2020, \textbf{20}, 4403--4409\relax
\mciteBstWouldAddEndPuncttrue
\mciteSetBstMidEndSepPunct{\mcitedefaultmidpunct}
{\mcitedefaultendpunct}{\mcitedefaultseppunct}\relax
\EndOfBibitem
\bibitem[Bera \emph{et~al.}(2017)Bera, Deb, Kathirvel, Bera, Thapa, and
  Saha]{bera2017flexible}
A.~Bera, K.~Deb, V.~Kathirvel, T.~Bera, R.~Thapa and B.~Saha, \emph{Appl. Surf.
  Sci.}, 2017, \textbf{418}, 264--269\relax
\mciteBstWouldAddEndPuncttrue
\mciteSetBstMidEndSepPunct{\mcitedefaultmidpunct}
{\mcitedefaultendpunct}{\mcitedefaultseppunct}\relax
\EndOfBibitem
\bibitem[Cheng \emph{et~al.}(2012 DOI: 10.1155/2012/593245)Cheng, Yu, Xing, and
  Yang]{cheng2012enhanced}
X.~Cheng, X.~Yu, Z.~Xing and L.~Yang, \emph{Int. J. Photoenergy}, 2012 DOI:
  10.1155/2012/593245\relax
\mciteBstWouldAddEndPuncttrue
\mciteSetBstMidEndSepPunct{\mcitedefaultmidpunct}
{\mcitedefaultendpunct}{\mcitedefaultseppunct}\relax
\EndOfBibitem
\bibitem[Ravi \emph{et~al.}(2017)Ravi, Santra, Joshi, Chugh, Singh, Rensmo,
  Ghosh, and Nag]{ravi2017origin}
V.~K. Ravi, P.~K. Santra, N.~Joshi, J.~Chugh, S.~K. Singh, H.~Rensmo, P.~Ghosh
  and A.~Nag, \emph{J. Phys. Chem. Lett}, 2017, \textbf{8}, 4988--4994\relax
\mciteBstWouldAddEndPuncttrue
\mciteSetBstMidEndSepPunct{\mcitedefaultmidpunct}
{\mcitedefaultendpunct}{\mcitedefaultseppunct}\relax
\EndOfBibitem
\bibitem[Tauc \emph{et~al.}(1966)Tauc, Grigorovici, and Vancu]{tauc1966optical}
J.~Tauc, R.~Grigorovici and A.~Vancu, \emph{Phys. Status Solidi B}, 1966,
  \textbf{15}, 627--637\relax
\mciteBstWouldAddEndPuncttrue
\mciteSetBstMidEndSepPunct{\mcitedefaultmidpunct}
{\mcitedefaultendpunct}{\mcitedefaultseppunct}\relax
\EndOfBibitem
\bibitem[Zhang \emph{et~al.}(2019)Zhang, Su, Lin, Liu, Chang, and
  Hao]{zhang2019disappeared}
J.~Zhang, J.~Su, Z.~Lin, M.~Liu, J.~Chang and Y.~Hao, \emph{Appl. Phys. Lett.},
  2019, \textbf{114}, 181902\relax
\mciteBstWouldAddEndPuncttrue
\mciteSetBstMidEndSepPunct{\mcitedefaultmidpunct}
{\mcitedefaultendpunct}{\mcitedefaultseppunct}\relax
\EndOfBibitem
\bibitem[Koscher \emph{et~al.}(2017)Koscher, Swabeck, Bronstein, and
  Alivisatos]{koscher2017essentially}
B.~A. Koscher, J.~K. Swabeck, N.~D. Bronstein and A.~P. Alivisatos, \emph{J.
  Am. Chem. Soc.}, 2017, \textbf{139}, 6566--6569\relax
\mciteBstWouldAddEndPuncttrue
\mciteSetBstMidEndSepPunct{\mcitedefaultmidpunct}
{\mcitedefaultendpunct}{\mcitedefaultseppunct}\relax
\EndOfBibitem
\bibitem[Ghosh \emph{et~al.}(2019)Ghosh, Jana, Sain, Ghosh, and
  Patra]{ghosh2019influence}
G.~Ghosh, B.~Jana, S.~Sain, A.~Ghosh and A.~Patra, \emph{Phys. Chem. Chem.
  Phys.}, 2019, \textbf{21}, 19318--19326\relax
\mciteBstWouldAddEndPuncttrue
\mciteSetBstMidEndSepPunct{\mcitedefaultmidpunct}
{\mcitedefaultendpunct}{\mcitedefaultseppunct}\relax
\EndOfBibitem
\bibitem[Tama \emph{et~al.}(2019)Tama, Das, Dutta, Bhuyan, Islam, and
  Basith]{tama2019mos}
A.~M. Tama, S.~Das, S.~Dutta, M.~D.~I. Bhuyan, M.~N. Islam and M.~A. Basith,
  \emph{RSC Adv.}, 2019, \textbf{9}, 40357--40367\relax
\mciteBstWouldAddEndPuncttrue
\mciteSetBstMidEndSepPunct{\mcitedefaultmidpunct}
{\mcitedefaultendpunct}{\mcitedefaultseppunct}\relax
\EndOfBibitem
\bibitem[Watanabe \emph{et~al.}(1977)Watanabe, Takizawa, and
  Honda]{watanabe1977photocatalysis}
T.~Watanabe, T.~Takizawa and K.~Honda, \emph{J. Phys. Chem.}, 1977,
  \textbf{81}, 1845--1851\relax
\mciteBstWouldAddEndPuncttrue
\mciteSetBstMidEndSepPunct{\mcitedefaultmidpunct}
{\mcitedefaultendpunct}{\mcitedefaultseppunct}\relax
\EndOfBibitem
\bibitem[Hu \emph{et~al.}(2006)Hu, Mohamood, Ma, Chen, and
  Zhao]{hu2006oxidative}
X.~Hu, T.~Mohamood, W.~Ma, C.~Chen and J.~Zhao, \emph{J. Phys. Chem. B}, 2006,
  \textbf{110}, 26012--26018\relax
\mciteBstWouldAddEndPuncttrue
\mciteSetBstMidEndSepPunct{\mcitedefaultmidpunct}
{\mcitedefaultendpunct}{\mcitedefaultseppunct}\relax
\EndOfBibitem
\bibitem[Segall \emph{et~al.}(2002)Segall, Lindan, Probert, Pickard, Hasnip,
  Clark, and Payne]{segall2002first}
M.~D. Segall, P.~J.~D. Lindan, M.~J. Probert, C.~J. Pickard, P.~J. Hasnip,
  S.~J. Clark and M.~C. Payne, \emph{J. Phys.: Condens. Matter}, 2002,
  \textbf{14}, 2717\relax
\mciteBstWouldAddEndPuncttrue
\mciteSetBstMidEndSepPunct{\mcitedefaultmidpunct}
{\mcitedefaultendpunct}{\mcitedefaultseppunct}\relax
\EndOfBibitem
\bibitem[Perdew \emph{et~al.}(1996)Perdew, Burke, and
  Ernzerhof]{perdew1996generalized}
J.~P. Perdew, K.~Burke and M.~Ernzerhof, \emph{Phys. Rev. Lett.}, 1996,
  \textbf{77}, 3865\relax
\mciteBstWouldAddEndPuncttrue
\mciteSetBstMidEndSepPunct{\mcitedefaultmidpunct}
{\mcitedefaultendpunct}{\mcitedefaultseppunct}\relax
\EndOfBibitem
\bibitem[Fischer and Almlof(1992)]{fischer1992general}
T.~H. Fischer and J.~Almlof, \emph{J. Phys. Chem.}, 1992, \textbf{96},
  9768--9774\relax
\mciteBstWouldAddEndPuncttrue
\mciteSetBstMidEndSepPunct{\mcitedefaultmidpunct}
{\mcitedefaultendpunct}{\mcitedefaultseppunct}\relax
\EndOfBibitem
\bibitem[Monkhorst and Pack(1976)]{monkhorst1976special}
H.~J. Monkhorst and J.~D. Pack, \emph{Phys. Rev. B}, 1976, \textbf{13},
  5188\relax
\mciteBstWouldAddEndPuncttrue
\mciteSetBstMidEndSepPunct{\mcitedefaultmidpunct}
{\mcitedefaultendpunct}{\mcitedefaultseppunct}\relax
\EndOfBibitem
\bibitem[Shenton \emph{et~al.}(2017)Shenton, Bowler, and
  Cheah]{shenton2017effects}
J.~K. Shenton, D.~R. Bowler and W.~L. Cheah, \emph{J. Phys.: Condens. Matter},
  2017, \textbf{29}, 445501\relax
\mciteBstWouldAddEndPuncttrue
\mciteSetBstMidEndSepPunct{\mcitedefaultmidpunct}
{\mcitedefaultendpunct}{\mcitedefaultseppunct}\relax
\EndOfBibitem
\bibitem[Basith \emph{et~al.}(2018)Basith, Yesmin, and Hossain]{basith2018low}
M.~A. Basith, N.~Yesmin and R.~Hossain, \emph{RSC Adv.}, 2018, \textbf{8},
  29613--29627\relax
\mciteBstWouldAddEndPuncttrue
\mciteSetBstMidEndSepPunct{\mcitedefaultmidpunct}
{\mcitedefaultendpunct}{\mcitedefaultseppunct}\relax
\EndOfBibitem
\bibitem[Deng \emph{et~al.}(2019)Deng, Zhao, Zhao, and
  Cai]{deng2019theoretical}
X.-Z. Deng, Q.-Q. Zhao, Y.-Q. Zhao and M.-Q. Cai, \emph{Curr. Appl Phys.},
  2019, \textbf{19}, 279--284\relax
\mciteBstWouldAddEndPuncttrue
\mciteSetBstMidEndSepPunct{\mcitedefaultmidpunct}
{\mcitedefaultendpunct}{\mcitedefaultseppunct}\relax
\EndOfBibitem
\bibitem[Peng \emph{et~al.}(2016)Peng, Zhang, Shao, Xu, Zhang, and
  Zhu]{peng2016electronic}
B.~Peng, H.~Zhang, H.~Shao, Y.~Xu, R.~Zhang and H.~Zhu, \emph{J. Mater. Chem.
  C}, 2016, \textbf{4}, 3592--3598\relax
\mciteBstWouldAddEndPuncttrue
\mciteSetBstMidEndSepPunct{\mcitedefaultmidpunct}
{\mcitedefaultendpunct}{\mcitedefaultseppunct}\relax
\EndOfBibitem
\bibitem[Bredas(2014)]{bredas2014mind}
J.-L. Bredas, \emph{Mater. Horiz.}, 2014, \textbf{1}, 17--19\relax
\mciteBstWouldAddEndPuncttrue
\mciteSetBstMidEndSepPunct{\mcitedefaultmidpunct}
{\mcitedefaultendpunct}{\mcitedefaultseppunct}\relax
\EndOfBibitem
\bibitem[Yu \emph{et~al.}(2018)Yu, Ma, Zhao, Liu, and Cai]{yu2018surface}
Z.-L. Yu, Q.-R. Ma, Y.-Q. Zhao, B.~Liu and M.-Q. Cai, \emph{J. Phys. Chem. C},
  2018, \textbf{122}, 9275--9282\relax
\mciteBstWouldAddEndPuncttrue
\mciteSetBstMidEndSepPunct{\mcitedefaultmidpunct}
{\mcitedefaultendpunct}{\mcitedefaultseppunct}\relax
\EndOfBibitem
\bibitem[Zhang \emph{et~al.}(2012)Zhang, Liu, and Zhou]{zhang2012towards}
H.~Zhang, L.~Liu and Z.~Zhou, \emph{Phys. Chem. Chem. Phys.}, 2012,
  \textbf{14}, 1286--1292\relax
\mciteBstWouldAddEndPuncttrue
\mciteSetBstMidEndSepPunct{\mcitedefaultmidpunct}
{\mcitedefaultendpunct}{\mcitedefaultseppunct}\relax
\EndOfBibitem
\bibitem[Zhang \emph{et~al.}(2015)Zhang, Yu, Liu, and
  Liu]{zhang2015illustration}
J.~Zhang, W.~Yu, J.~Liu and B.~Liu, \emph{Appl. Surf. Sci.}, 2015,
  \textbf{358}, 457--462\relax
\mciteBstWouldAddEndPuncttrue
\mciteSetBstMidEndSepPunct{\mcitedefaultmidpunct}
{\mcitedefaultendpunct}{\mcitedefaultseppunct}\relax
\EndOfBibitem
\bibitem[Phillips(1968)]{phillips1968covalent}
J.~C. Phillips, \emph{Phys. Rev.}, 1968, \textbf{166}, 832\relax
\mciteBstWouldAddEndPuncttrue
\mciteSetBstMidEndSepPunct{\mcitedefaultmidpunct}
{\mcitedefaultendpunct}{\mcitedefaultseppunct}\relax
\EndOfBibitem
\bibitem[Dalpian \emph{et~al.}(2017)Dalpian, Liu, Stoumpos, Douvalis,
  Balasubramanian, Kanatzidis, and Zunger]{dalpian2017changes}
G.~M. Dalpian, Q.~Liu, C.~C. Stoumpos, A.~P. Douvalis, M.~Balasubramanian,
  M.~G. Kanatzidis and A.~Zunger, \emph{Phys. Rev. Mater.}, 2017, \textbf{1},
  025401\relax
\mciteBstWouldAddEndPuncttrue
\mciteSetBstMidEndSepPunct{\mcitedefaultmidpunct}
{\mcitedefaultendpunct}{\mcitedefaultseppunct}\relax
\EndOfBibitem
\end{mcitethebibliography}
